%% file: main.tex
\documentclass[a4paper,11pt]{article}
\usepackage[utf8]{inputenc}

\usepackage{graphicx}
\usepackage[version=3]{mhchem}
\usepackage{subcaption}
\usepackage{float}
\usepackage{jinstpub}
\usepackage{color, colortbl}
\usepackage{amsmath}
\usepackage{appendix}
\usepackage{footmisc}
\definecolor{Gray}{gray}{0.9}

\usepackage[colorinlistoftodos]{todonotes}

\begin{document}
\title{The Electric Field Dependence of Single Electron Emission in the PIXeY Two-Phase Xenon Detector}
\author[a, b, e]{E. Bodnia,}
\author[a,b,c]{E. P. Bernard,}%
\author[a]{A.~Biekert,}%
\author[a,b,c]{E.~M.~Boulton,}%
\author[c]{S.~B.~Cahn,}%
\author[,d]{N.~Destefano,$^1$}\footnote{The MITRE Corporation, 202 Burlington Rd, Bedford, MA 01730}
\author[,c]{B.~N.~V.~Edwards,$^2$}\footnote{IBM Research, STFC Daresbury Laboratory, Warrington, WA4 4AD, UK}
\author[c,d]{M.~Gai,}
\author[,a,b,c]{M.~Horn,$^3$}\footnote{Sanford Underground Research Facility, 630 E. Summit Street, Lead, SD 57754}
\author[,c]{N.A.~Larsen,$^4$}\footnote{Kavli Institute for Cosmological Physics, University of Chicago, 5640 Ellis Ave, Chicago, IL 60637}
\author[b]{Q.~Riffard,}%
\author[,c]{B.~Tennyson,$^1$}%%
\author[a]{V.~Velan,}%
\author[,c]{C.~Wahl,$^6$}\footnote{H3D, Inc., 812 Avis Dr, Ann Arbor, MI 48108}%
\author[a,b,c]{and D.~N.~McKinsey}%

\affiliation[a]{Department of Physics, University of California Berkeley, 366 Physics North, Berkeley, CA 94720, USA}
\affiliation[b]{Lawrence Berkeley National Laboratory, 1 Cyclotron Rd, Berkeley, CA 94720, USA}
\affiliation[c]{Yale University, Department of Physics, 217 Prospect St., New Haven, CT 06511, USA}
\affiliation[d]{University of Connecticut, LNS at Avery Point, 1084 Shennecossett Road, Groton, CT 06340, USA}
\affiliation[e]{Department of Physics, University of California Santa Barbara, Broida Hall,  Santa Barbara, CA 93106, USA}

\emailAdd{ebodnia@ucsb.edu}


\abstract{Dual phase xenon detectors are widely used in experimental searches for galactic dark matter particles. The origin of single electron backgrounds following prompt scintillation and proportional scintillation signals in these detectors is not fully understood, although there has been progress in recent years. In this paper, we describe single electron backgrounds in \ce{^{83m}_{}Kr} calibration events and their correlation with drift and extraction fields, using the Particle Identification in Xenon at Yale (PIXeY) dual-phase xenon time projection chamber. The single electron background induced by the Fowler-Nordheim (FN) effect is measured, and its electric field dependence is quantified. The photoionization of grids and impurities by prompt scintillation and proportional scintillation also contributes to the single electron background.
}

\maketitle

\flushbottom

\section{Introduction}

	Dual phase time projection chambers (TPCs) are widely used in experimental searches for galactic dark matter particles scattering on target nuclei or electrons. Energy depositions in liquid xenon (LXe) produce signals via prompt scintillation light and ionization charge. The prompt scintillation light is detected in photomultiplier tubes (PMTs) at the top and bottom of the detector, producing a signal known as S1. Liberated electrons drift upwards in an applied electric field and generate delayed electroluminescence signal, called S2, in the gaseous xenon (GXe)~\cite{luxKr, Boulton}. Several experiments have observed electron backgrounds on both microsecond and millisecond time scales following S2 pulses, which are referred to as "single electron trains", "small S2 signals", "emission currents",  or just "single electrons"~\cite{Santos, Angle, Sorensen}. This background limits sensitivity to low-mass and MeV-scale hidden sector dark matter searches and complicates the classification of low-energy events~\cite{Sorensen}. Therefore, it is important to study single electrons and identify their origins. There are many mechanisms that may account for single electron production: metal wire field emission, metal grid photoelectrons induced by S1 and S2 ultraviolet (VUV) photons, release of trapped thermalized electrons from the liquid surface, photoionization of impurities from the liquid xenon active volume or liquid-gas interface, and motion of the liquid surface.
	
	   The ZEPLIN-III \cite{Santos}, ZEPLIN-II \cite{Edwards}, and XENON \cite{Aprile} collaborations studied photon-induced and spontaneous electron emission.  Single electrons were categorized according to their drift time (i.e. depth) measurements and studied as a function of electron lifetime. Emission with electron drift time near the maximum available drift time was labeled as `S1-induced emission' and understood to originate from the cathode grid. Single electrons with shorter drift time were hypothesized to originate from S1-photoionized impurities in the liquid xenon and also classified asS1-induced emission. Finally, spontaneous emission was the category given to single electrons that occur later than the maximum electron drift time in the TPC. The possible explanations of such emission were stainless-steel field emission and delayed emission of thermal electrons trapped at the surface \cite{Edwards, Santos}. XENON100 also examined single electron signals and reported a positive correlation between the single electron rate and the $\rm  O_2$ impurity level, and they inferred that the observed small S2 signals are related to the photoionization of $\rm O_2^{-}$ impurities ~\cite{Aprile}. 
      % 
    % The cathode grids in \cite{Bailey, Tomas} studies were electropolished which significantly reduced protruding defects, so the theory of FN current originating at small protruding filaments was excluded.
    
    Spurious backgrounds from the cathodic grids were studied in \cite{Bailey}. The cathodic metal wires and grids emit electrons once an external electric field is applied, and the origin of this phenomenon can be described within the framework of Fowler-Nordheim (FN) theories. One potential explanation is the projection model, which implies that the field emission occurs at localized areas on the cathode surface. The electron potential energy just outside the metal surface decreases linearly due to the applied field, and the step potential at the metal-xenon interface is rounded due to the image charge seen by an electron in LXe. This allows conduction electrons near the Fermi level to tunnel into the lower-potential regions of the surrounding LXe, resulting in an emission current ranging between $1- 1000$ nA ~\cite{Nordheim, grids, Alpert}.

    Other field emission studies were done in preparation for the LUX-ZEPLIN (LZ) experiment, using different wire samples \citep{grids}. The measured emission was classified according to the emitter's rate and the range of fields over which emission  occurred: \textit{extended emitters} persisted over a wide range of fields (higher than 1 kV/cm); \textit{impulsive emitters} oscillated in time; and \textit{faint emission} occurred at a low, steady field-independent rate. A few conclusions were made based on these studies. First, no intrinsic threshold was revealed that would lead to electric breakdown of cathodic surfaces in LXe, even at field strengths higher than 160 kV/cm. Second, the macroscopic field on the wire surface appeared not to be the key parameter explaining most phenomena, so additional explanations of the high voltage breakdowns were suggested, such as the Malter effect \cite{Malter} and a protrusion model characterized by the enhancement factor $\beta$. The Malter effect could be associated with Xe$^+$ ions accumulating on the cathode, resulting in simultaneous electron and photon production. The protrusion model explains the emission currents as due to impurities embedded or accumulated in the metal, or by mechanical defects on the grids resulting in sharp tips protruding from the metal surface. In the vicinity of these tips, the electric field can be as high as $10^9$ V/m ~\citep{Alpert, Noer, Santos}. The expected field threshold to produce charge multiplication and light emission in LXe is expected to be $700$ kV/cm field on wires \citep{Aprile:2014}, so if one considers the filaments model, the enhancement factor would be limited to the $10-20$ range. However, the protrusion model predicts $\beta \approx 100-1000$ to observe a significant emission current. Although the authors did not favor the filament hypothesis, they also did not exclude this idea. The authors also consider the possibility that small stressed areas may tolerate higher electric fields without dielectric breakdown \citep{MicroBooNE,XeBrA}, thus allowing higher values of $\beta$ without breakdown for small filaments or mechanical defects. 
 	
 	More research on single electron backgrounds has been done by the Large Underground Xenon (LUX) collaboration for a single drift field of 180~V/cm applied across the TPC ~\citep{Jingke}. According to this analysis, single electrons can be clustered and originate from instrumental effects, capture of electrons by impurities and subsequent photoionization, electron trapping, liquid surface fluctuations, and grid emission. The delayed single electrons are investigated for a 0--3 ms time window following the S2 signal for \ce{^{83m}_{}Kr} calibration events, and for a 3--1000 ms time window for low-background WIMP-search data. The emission of electron clusters (e-bursts) decays exponentially. Moreover, the single electron rate following interactions in the bottom of the TPC is higher compared to events in the top of the TPC. The single electron rate exhibits similar time dependence for both top and bottom TPC events, preserves X-Y position correlation, and follows a power-law time dependence.  
    
	The thermalization trapping and release of single electrons at the liquid-gas interface hypothesis was suggested and studied in \cite{Sorensen, Akimov}. The delayed single electron noise in~\cite{Sorensen} is split into two distinct exponential components, obtained by fitting a linear function to the logarithm of the sum of the observed events over two separate region: $\tau_1$ is fit between 40--80 $\mu s$ after the S2, and $\tau_2$ is fit between 500--1000 $\mu s$ after the S2. The components decaying with lifetimes $\tau_1 $ and $\tau_2$ are called \textit{fast} and \textit{slow} components, respectively, and studied as a function of the applied electric field. The electron train fast component amplitude, expressed as a percentage of the original number of electrons in the ionizing event, decreases with increased field, while the slow component increases linearly with the electric field. The $\tau_1$ trend is explained by thermalized, unextracted primary electrons; while the origins of $\tau_2$ are hypothesized to be due to ionization of negative ion impurities by VUV photons coming from the weakly fluorescent Polytetrafluoroethylene (PTFE) walls inside the detector \cite{Sorensen}. A similar conclusion on single electron emission is made in ~\cite{Akimov}. The single electron emission is also observed on a millisecond time scale and explained to be due to electron trapping near the liquid surface and electron capture by electronegative impurities. The single electron rate after the electroluminescent signals following S2 corresponding to single-vertex interactions of gammas or neutrons in LXe is fit to a power law function, which is compatible with LUX observations~\citep{Jingke}.

\section{The PIXeY Detector}

\subsection{Setup}
\begin{figure}[H]
\begin{center}
\includegraphics[width=0.8\linewidth]{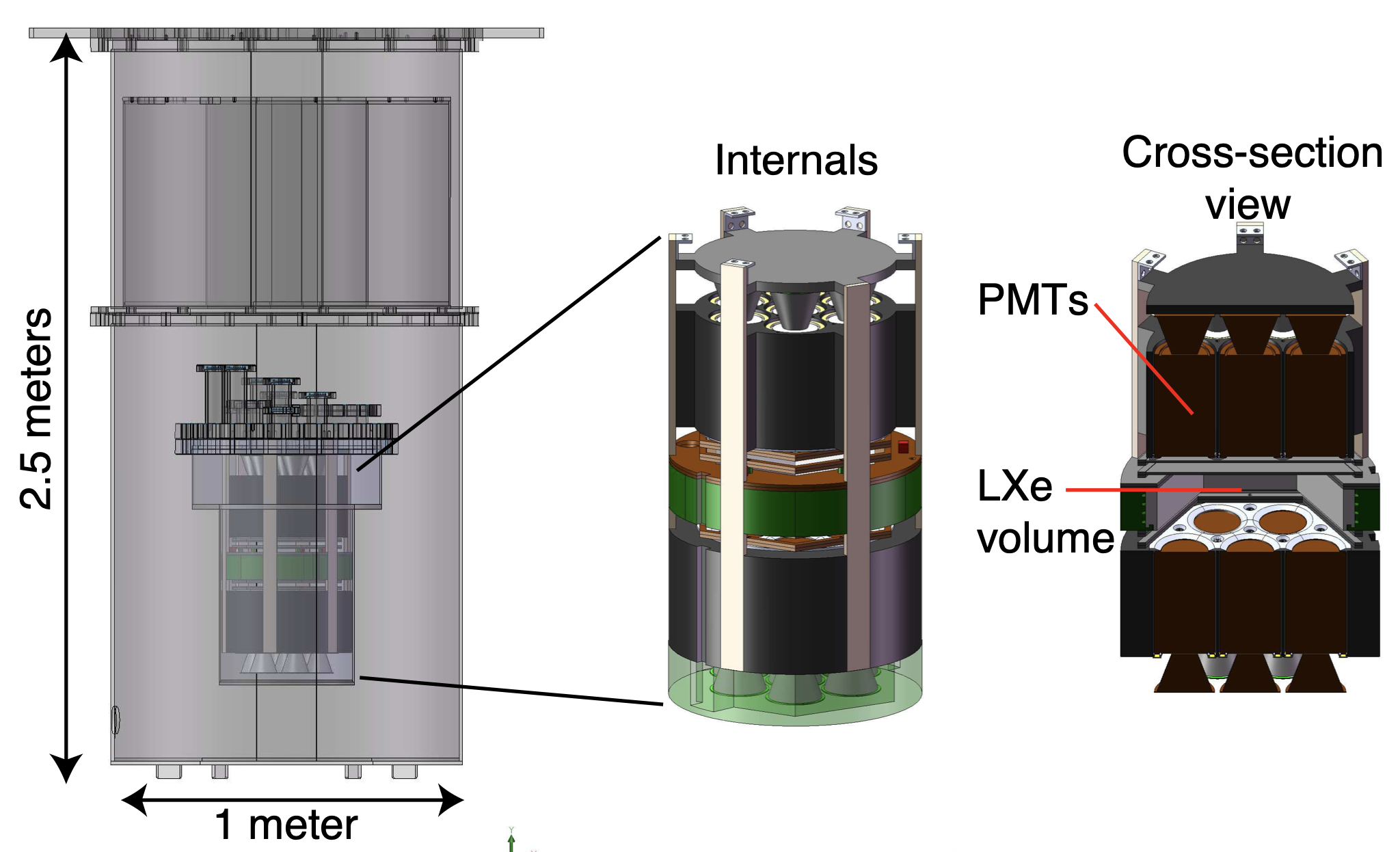}
\caption{A schematic of the PIXeY detector.}
\label{fig:1}
\end{center}
\end{figure}

The PIXeY detector is a hexagonal xenon TPC holding 3 kg of liquid xenon in its active volume, as shown in Fig.~\ref{fig:1}.  The main goals of this detector were to optimize energy resolution and background rejection by studying the effects of independently controlling the drift and electron-extraction fields. Also, the detector has been used to study electron extraction efficiency for the range of electric fields from 2.4 to 7.1 kV/cm in the liquid xenon, corresponding to 4.5 to 13.1 kV/cm in the gaseous xenon  ~\cite{edwards_eff} and the scintillation and ionization response of \ce{^{83m}_{}Kr} and \ce{^{37}_{}Ar} ~\cite{Boulton, Anisha} for a variety of electric field conditions. 

\begin{figure}[H]
\begin{center}
\includegraphics[width=0.8\linewidth]{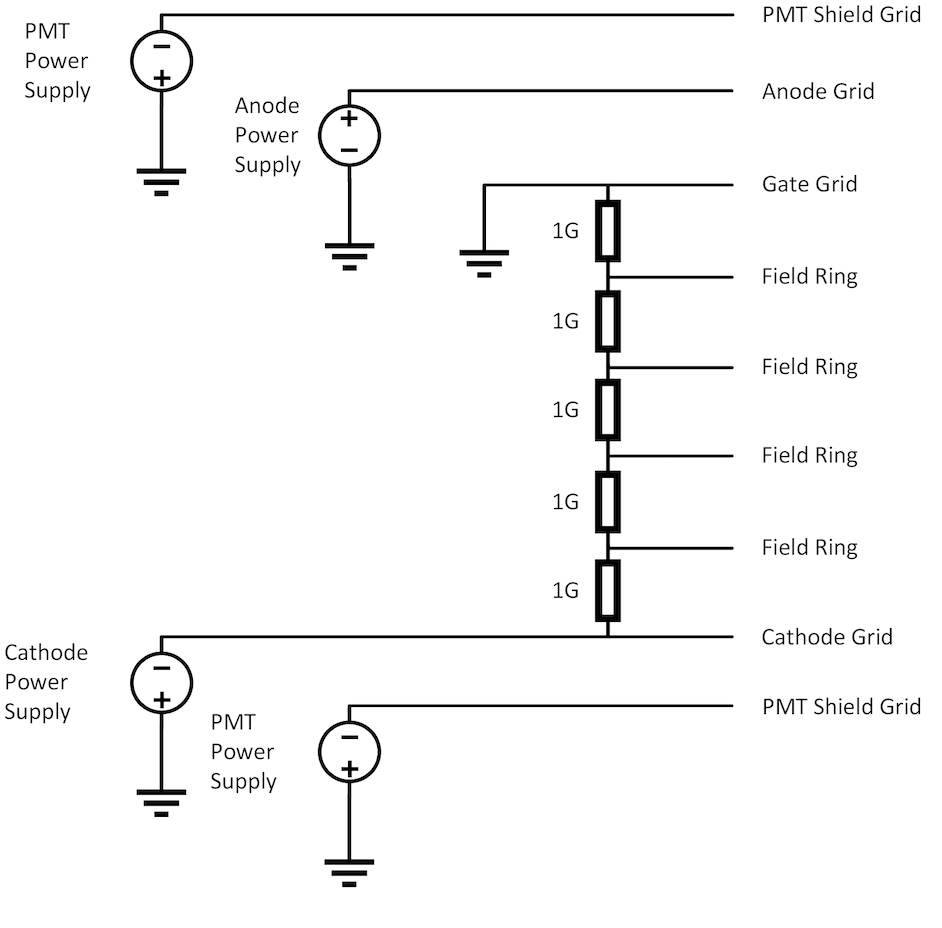}
\caption{A sketch of the PIXeY circuit. The anode and cathode power supplies are connected directly to the anode and cathode correspondingly. The voltages of the top and bottom shield grids are set to match the PMT cathode voltages to about -1 kV.  The springs are 1 GOhm resistors bridging the gaps for the four field shaping rings (FR) between the cathode and the gate.}
\label{fig:circuit}
\end{center}
\end{figure}

The hexagonal volume has side lengths of 9.2~cm and a cathode-to-gate drift length of 5.1~cm. It has two arrays of seven Hamamatsu R8778 photomultiplier tubes (PMTs) with 33\% quantum efficiency, located on the top and bottom of the TPC. The PMT signal undergoes an eightfold amplification and is digitized with a 12-bit ADC (CAEN V1720) at 250 MHz~\cite{edwards_eff}. The detector includes a gate grid that provides separate control of the electric fields in the liquid and gaseous regions of the detector. High voltage epoxy feedthroughs are installed with stable operating voltages of 5 kV on the anode and -10 kV on the cathode. The extraction field is formed by the gate and the anode grids, which are separated by 0.7 cm. The schematic of the PIXeY circuit is shown at Fig. ~\ref{fig:circuit}, where the anode and cathode power supply was connected directly to the anode and cathode correspondingly, having no limiting resistor in between. Additionally, there were four field shaping rings between the cathode and the gate, forming five gaps, where each gap was bridged by a single 1 Gohm resistor. The data acquisition system is triggered on S2 pulses, which generally occur $85 \pm 5$ $\mu$s into the waveform. For the analysis, we used two categories of datasets. In the first category, the anode was fixed at 2.5 kV, while the cathode varied between 0.5--5 kV, giving a drift field in the range 0.098--0.98 kV/cm. In the second category, the cathode was fixed at 1 kV, while the anode voltage varied in the range 3.2--8.8 kV, corresponding to a liquid extraction field in the range 2.7--7.1 kV/cm. The anode voltage was dialed in on a power supply that needed to be calibrated, so the anode voltage used in the data analysis is given as (displayed anode voltage)$ \cdot 2.1 + 260$ V.

 PTFE walls are used for their high VUV reflectivity throughout the detector and for the surrounding plastic construction, including the parts that do not come into contact with the scintillation light~\cite{evan}. The electron lifetime was observed to be 500 $\mu$s or greater during the course of the runs. The electron signal magnitudes and widths are listed in Table 1.

\begin{table}[H]
\begin{center}
\begin{tabular}{lll}
\hline
\multicolumn{1}{|l|}{Liquid Extraction Field (kV/cm)} & \multicolumn{1}{l|}{Single Electron Size (phe)} & \multicolumn{1}{l|}{Single Electron Width ($\mu$s)} \\ \hline
\multicolumn{1}{|l|}{2.7}                      & \multicolumn{1}{l|}{20.44 $\pm$ 8.15}                      & \multicolumn{1}{l|}{0.87 $\pm$ 0.36}                       \\ \hline
\multicolumn{1}{|l|}{3.6}                      & \multicolumn{1}{l|}{29.22 $\pm$ 10.50}                      & \multicolumn{1}{l|}{0.78 $\pm$ 0.22}                       \\ \hline
\multicolumn{1}{|l|}{4.4}                      & \multicolumn{1}{l|}{36.74 $\pm$ 9.44}                      & \multicolumn{1}{l|}{0.71 $\pm$ 0.14}                       \\ \hline
\multicolumn{1}{|l|}{5.3}                      & \multicolumn{1}{l|}{47.79 $\pm$ 10.90}                      & \multicolumn{1}{l|}{0.60 $\pm$ 0.11}                       \\ \hline
\multicolumn{1}{|l|}{5.6}                      & \multicolumn{1}{l|}{52.35 $\pm$ 11.07}                      & \multicolumn{1}{l|}{0.58 $\pm$ 0.09}                       \\ \hline
\multicolumn{1}{|l|}{6.1}                      & \multicolumn{1}{l|}{57.81 $\pm$ 11.12}                      & \multicolumn{1}{l|}{0.54 $\pm$ 0.08}                       \\ \hline
\multicolumn{1}{|l|}{6.6}                      & \multicolumn{1}{l|}{61.68 $\pm$ 11.41}                      & \multicolumn{1}{l|}{0.51 $\pm$ 0.08}                       \\ \hline
\multicolumn{1}{|l|}{7.1}                      & \multicolumn{1}{l|}{67.50 $\pm$ 11.77}                      & \multicolumn{1}{l|}{0.48 $\pm$ 0.07}                       \\ \hline
\end{tabular}
\caption{The means and widths of the single electron signal distributions observed in \ce{^{83m}_{}Kr} events, as a function of extraction field.} 
\end{center}
\end{table}

\subsection{Signals in PIXeY}

\begin{figure}[H]
\begin{center}
\includegraphics[width=0.55\linewidth]{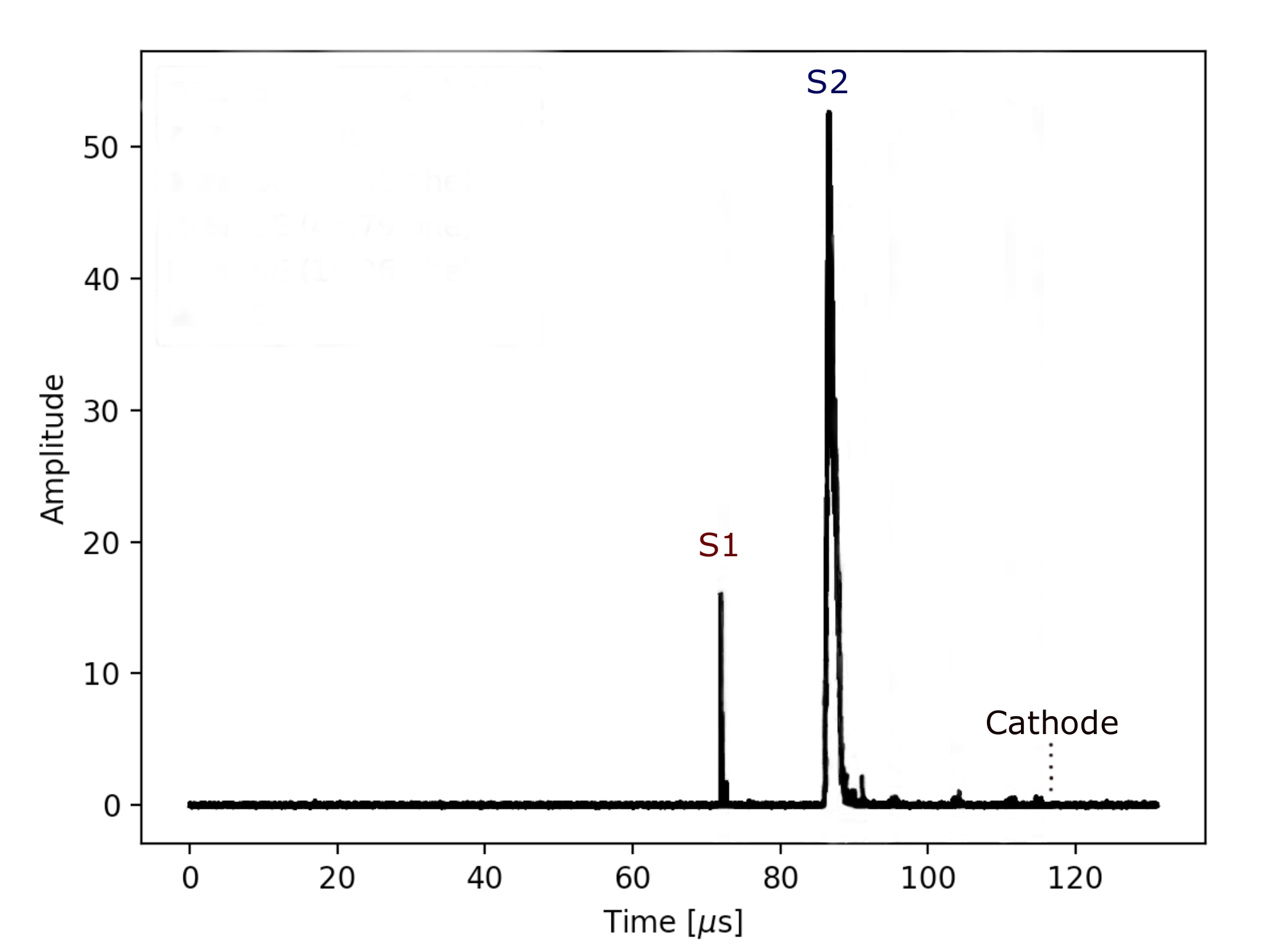}
\caption{An example of a single scatter waveform in the PIXeY detector \cite{edwards_eff}. S1 is the first scintillation signal which are the photons caused by \ce{^{83m}_{}Kr} isotopes decay to the ground state. S2 pulse is the ionization signal produced by the electrons resulted in that transition, hence follows shortly after the S1. The dashed line marked as "cathode" represent the maximum drift time in the TPC, which is a typical location of the cathode grids photoionization signal. In this figure, the S1a and S1b are close enough in time that they cannot be distinguished.}
\label{fig:wave}
\end{center}
\end{figure}

Data sets produced during \ce{^{83m}_{}Kr} calibrations are used for this analysis. \ce{^{83m}_{}Kr} is commonly introduced into low-background liquid xenon detectors, producing monoenergetic events to characterize the scintillation and ionization response. The \ce{^{83m}_{}Kr} generator was placed in a pressure differential in the GXe circulation path of the main chamber, dispersing \ce{^{83m}_{}Kr} atoms into liquid xenon resulting in 20 Hz of \ce{^{83m}_{}Kr} decays, uniformly distributed in the active volume~\cite{ks, Anisha}. The \ce{^{83m}_{}Kr} isotopes decay to the ground state in a two-step internal conversion, releasing conversion electrons with kinetic energies of 32.1 keV and 9.4 keV, respectively. The time between the steps is exponentially distributed with a half-life of 154 ns. The VUV light produced results in two S1 signals referred to S1a and S1b ~\cite{Anisha}. Since the decay time is much less than the microsecond variation in electron drift time due to diffusion, the two interactions occur at nearly the same time and the same location (Fig.~\ref{fig:wave}). Therefore, this produces a cluster of single electrons--around 750 electrons on average for the combined decay---which undergoes proportional scintillation in the gas layer and typically appears as a single S2 pulse~\cite{Larsen, Anisha}. 

\section{Selection of Single Electrons}

\begin{figure}[H]
\centering
\includegraphics[width=0.85\linewidth]{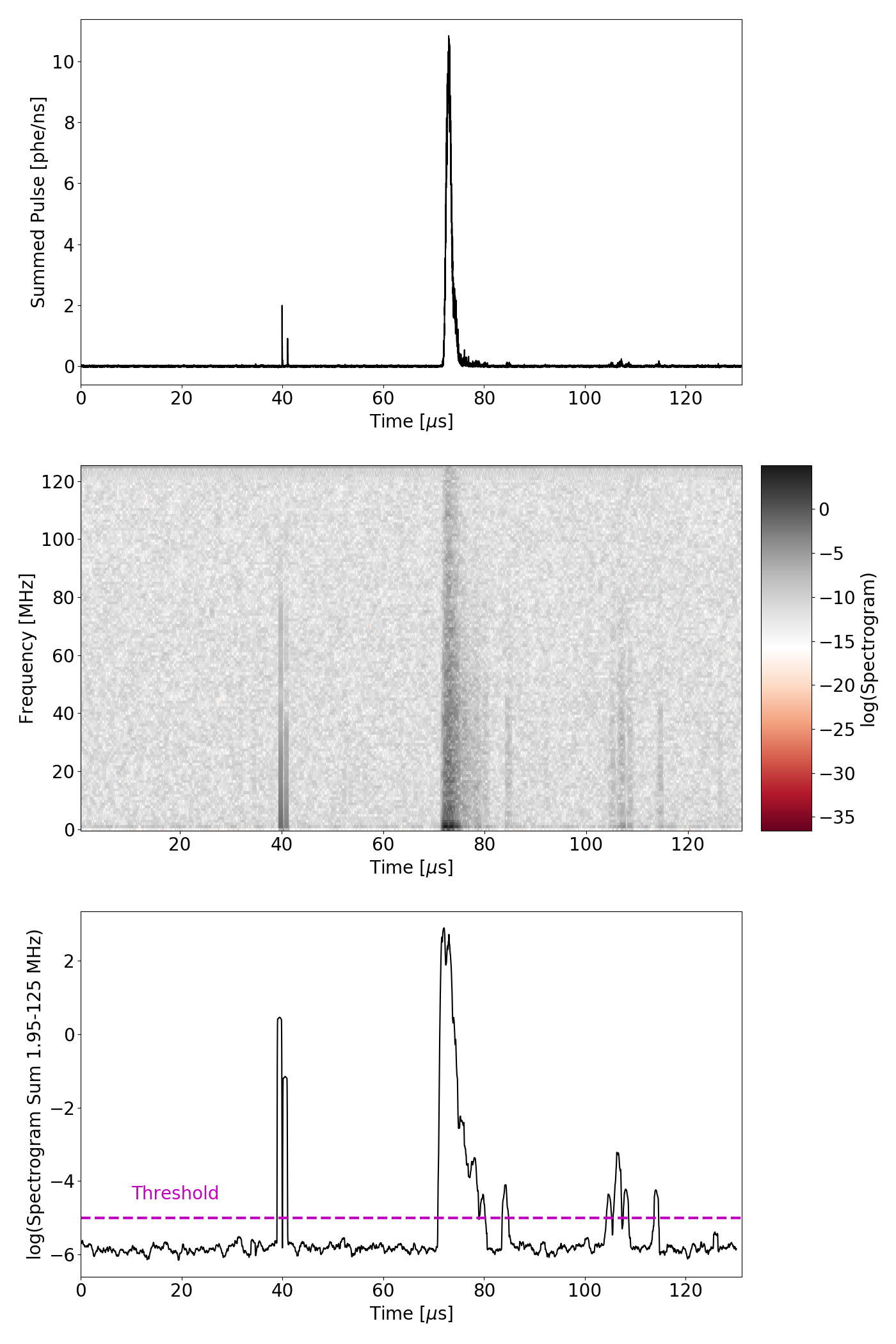}
\caption{The spectrogram procedure for a single \ce{^{83m}_{}Kr} event. (\textit{Top}) The summed waveform for a single event. The reader should note the S1a and S1b pulses at 40~$\mu s$, the S2 pulse at 70~$\mu s$, and several single electron pulses both immediately following the S2 and in the 100--120~$\mu s$ region. (\textit{Middle}) The spectrogram of this waveform in time vs. frequency space. There are obvious characteristics corresponding to the S1 and S2 pulses, but also note that the single electrons are more pronounced and separable than in the raw waveform. (\textit{Bottom}) The spectrogram summed across frequencies. This is the observable used to identify pulses, and the threshold for pulse selection is shown.}
\label{fig:spectro}
\end{figure}

\begin{figure}[H]
\begin{subfigure}{0.5\textwidth}
\includegraphics[width=1.1\linewidth]{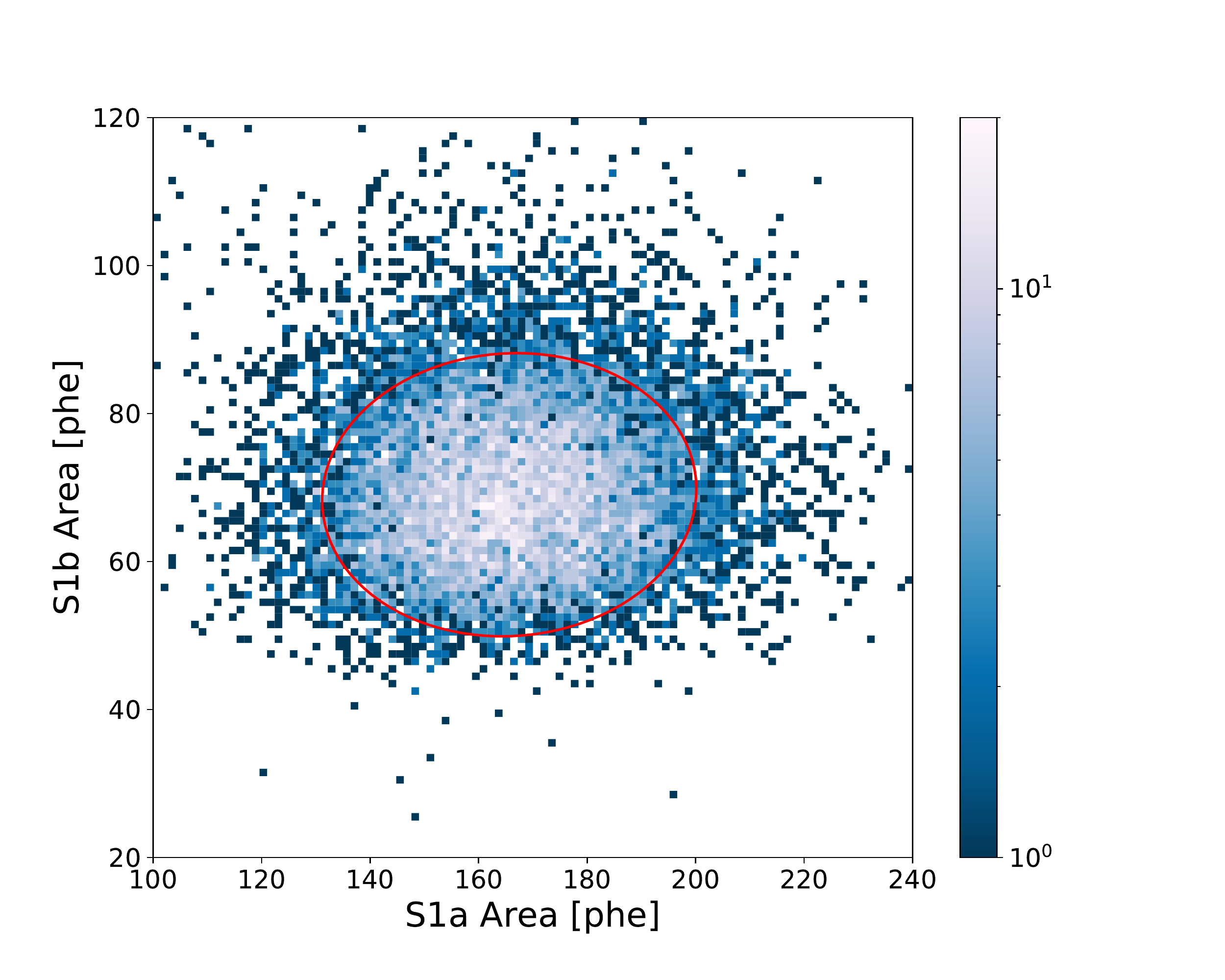} 

\end{subfigure}
\begin{subfigure}{0.5\textwidth}
\includegraphics[width=1.1\linewidth]{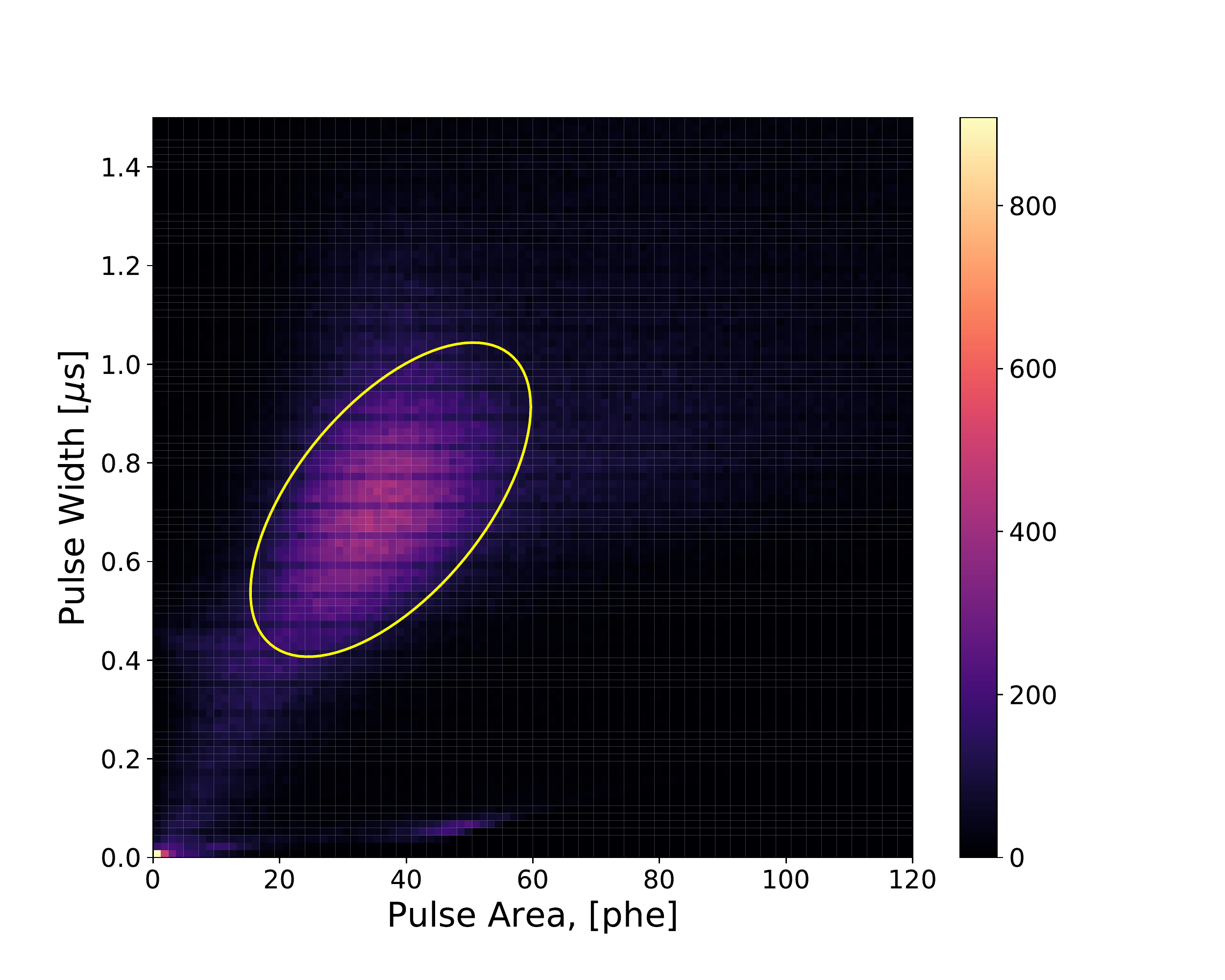}

\end{subfigure}

\caption{\textit{Left}: (S1a, S1b) distribution for \ce{^{83m}_{}}Kr events, given a 4.4 kV/cm extraction field and 0.2 kV/cm drift field. The red line shows the cut that was developed based on these variables. \textit{Right}: The distribution of single electron signals in events that are captured by the (S1a, S1b) cut. A 2D Gaussian cut based on pulse area and width is applied to select pulses.} 
\label{fig:2}
\end{figure}

To study the field-dependence of the single-electron (SE) background in a well-controlled data set, the first step is to find the pulses within the waveforms, classify them, and select only \ce{^{83m}_{}Kr} events, which have two primary scintillation (S1) pulses followed by one secondary scintillation (S2) pulse. Next, out of the selected \ce{^{83m}_{}Kr} events, we identify pulses by computing a spectrogram with a windowed Fourier transform. This allows us to decompose the waveform into a time-varying ensemble of frequencies, which is useful for selecting single electrons. A threshold on the power spectrum is chosen empirically to select regions of the waveform (i.e. pulses) where the power is maximized in the 1.95 to 125 MHz frequency range, where 125 MHz corresponds to the largest frequency signal that can be observed. Figure~\ref{fig:spectro} shows an example of this procedure. For the single electron pulses, we apply an additional 2D Gaussian cut on the resulting distribution of pulse area and height (Fig.~\ref{fig:2}, Right). To ensure quality in the \ce{^{83m}_{}Kr} events filter, a 2D Gaussian cut is applied to the S1a/S1b areas (Fig.~\ref{fig:2}, Left). Finally, we define the ``normalized single electrons'' as the summed SE area divided by the preceding pulse area (either an S1 or S2). For a given time period, the normalized SE \textit{rate} is found by dividing the normalized SE by the time scale under consideration. No fiducial cut is applied.

\section{Results}

In this section, we describe the behavior of single electrons during different time periods relative to the S1 and the S2 pulses within the selected \ce{^{83m}_{}Kr} events.

\subsection{Background on Long Time Scales}

The single-electron background rate at long time scales is the main concern for dark matter experiments. In order to assess this background, we investigate single electrons that occur before the S1a signal.

\begin{figure}[H]
\includegraphics[width=0.48\linewidth]{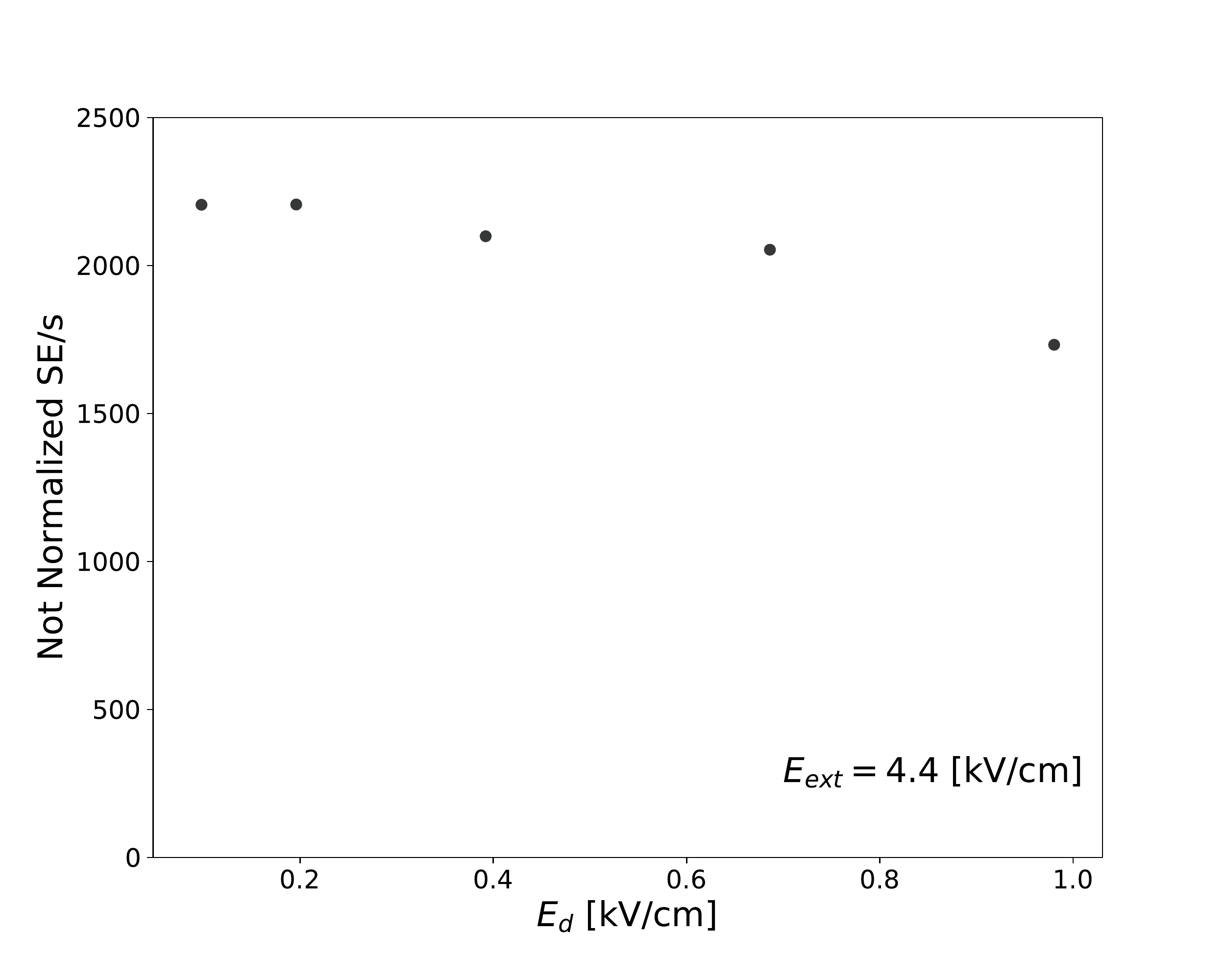} 
\includegraphics[width=0.48\linewidth]{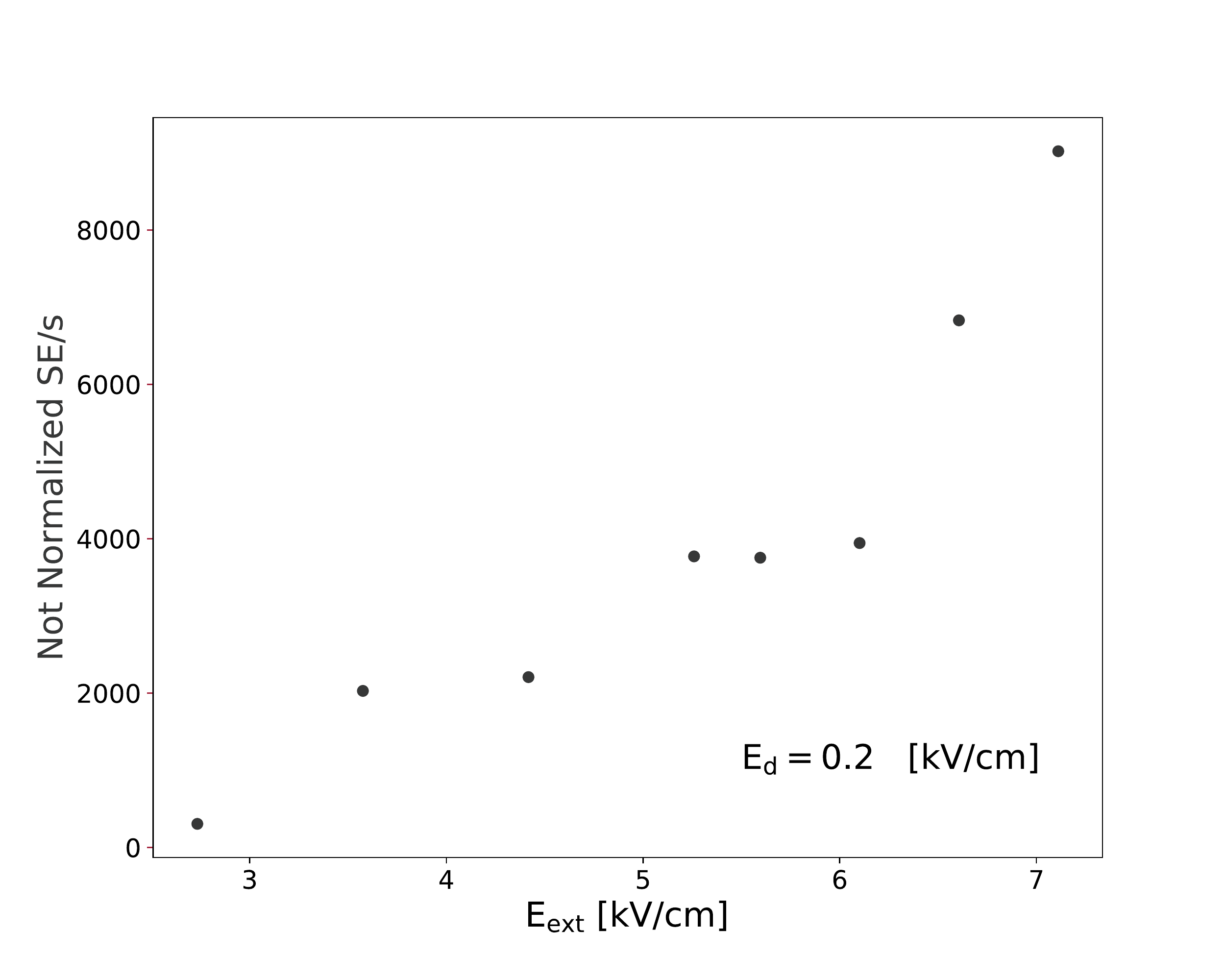}

\caption{The detected number of single electrons per second before the S1 pulse as a function of electric field.  One $\sigma$ error bars are shown for both figures, but are too small to be seen on the figure. \textit{Left:} SE rate for varied drift field and fixed extraction field of 4.4 kV/cm. \textit{Right:} SE rate for varied LXe extraction field and fixed drift field of 0.2 kV/cm.}
\label{fig:3}
\end{figure}

The rate of single electrons in this particular region is defined as the sum of the SE areas divided by the length of the time on which they are observed, divided by the $g_2$ gain. In turn, $g_2$ is defined as the number of detected photons per electron produced; it is dependent on the extraction field. %(timescale before the S1a is the difference between the length of the acquisition window and the start time of the S1a pulse).
The SE rate is somewhat suppressed as the drift field ($E_d$) grows, for an extraction field ($E_{ext}$) held at a constant value (Fig.~\ref{fig:3}, Left). The SE rate grows significantly with extraction field (Fig.~\ref{fig:3}, Right), scaling as the product of electron extraction efficiency and FN emission rate.

\begin{figure}[H]
\begin{subfigure}{0.5\textwidth}
\includegraphics[width=\linewidth]{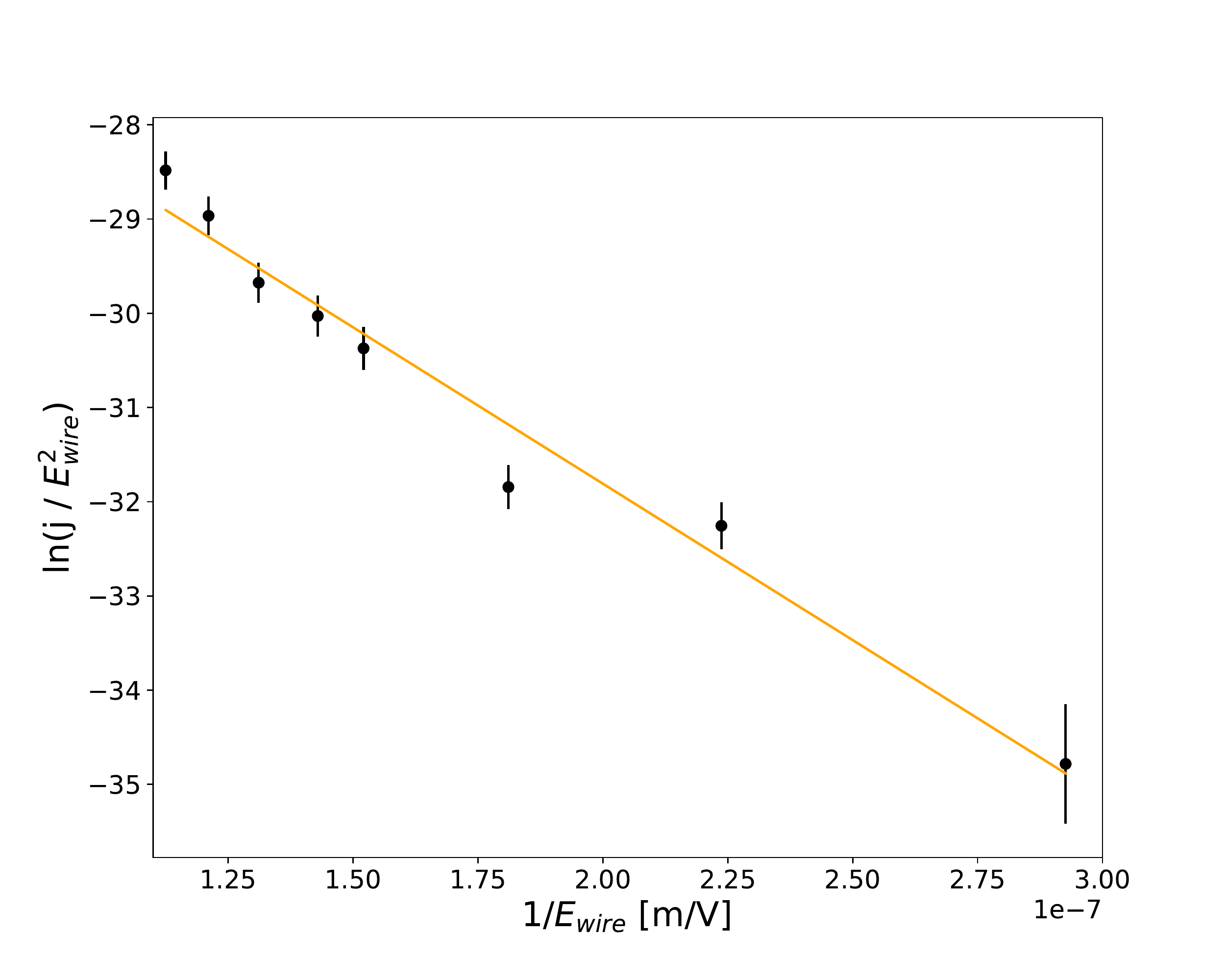}
\end{subfigure}
\begin{subfigure}{0.50\textwidth}
\includegraphics[width=\linewidth]{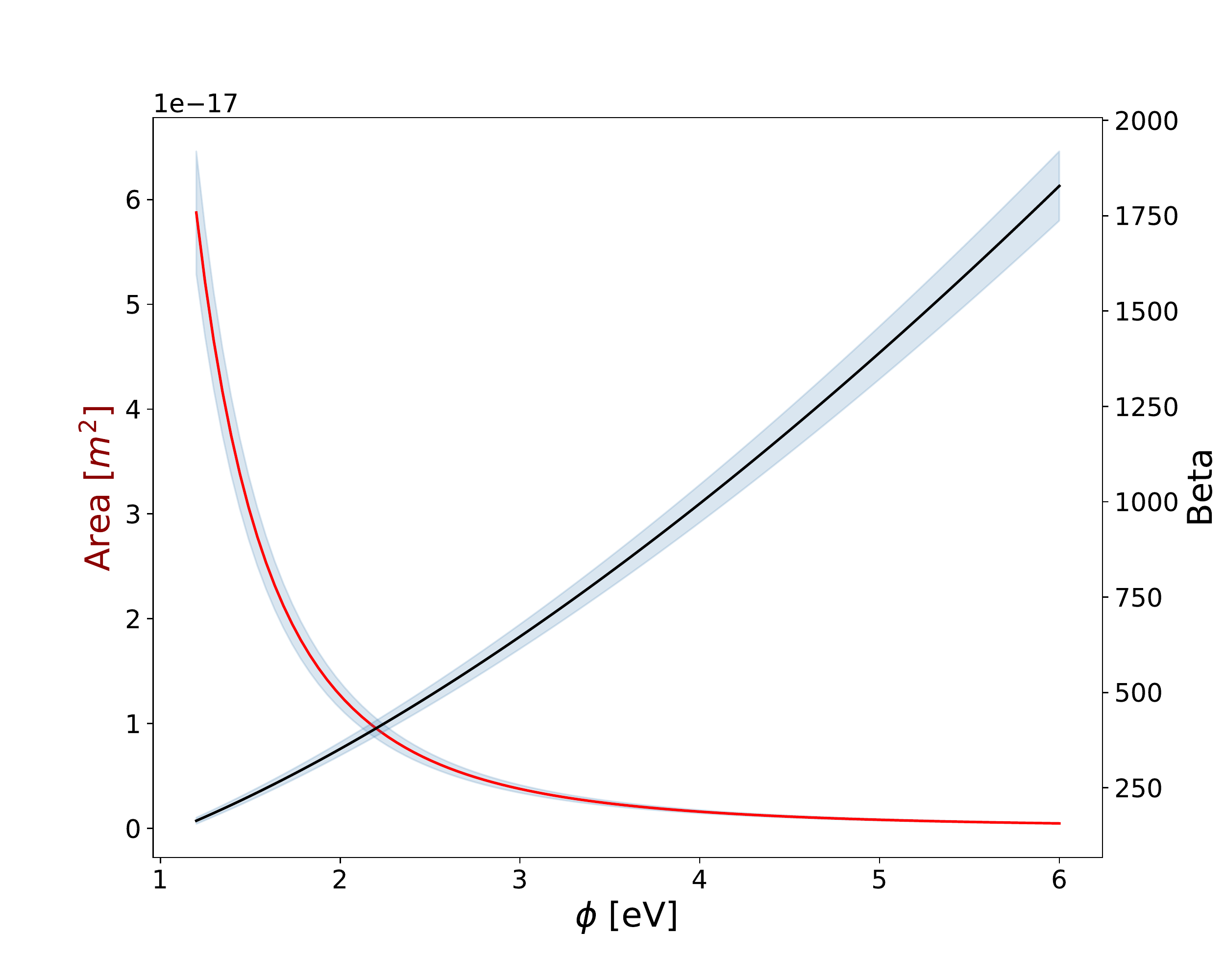}
\end{subfigure}
\caption{\textit{Left}: The Fowler-Nordheim plot for the emission current \textit{j}, in $\mathrm{\frac{A}{m^2}}, $ originating on the gate TPC wires. The emission current \textit{j} is corrected for electron extraction efficiency. The electric field $E_{wire}$ in $\mathrm{\frac{V}{m}}$ is calculated assuming a cylindrical wire approximation. \textit{Right:} The black curve is the dimensionless field enhancement factor $\beta$ as a function of the work function $\phi$ in eV. The red curve reflects the dependency of the emitter tip area on the wire work function. The error on each curve, indicated by the shaded region, is derived from the fit when extracting the $\beta$ factor.}
\label{fig:4}
\end{figure}

The  measured single electron background is observed to follow the Fowler-Nordheim equation~\ref{eq:4.1} (Fig.~\ref{fig:4}, Right), corrected for the electron extraction efficiency, the calculations and theory of which are described in ~\citep{edwards_eff, evan}:

\begin{equation}
      j(E) = \frac{A_0 E^2}{\phi t^2(y)} \cdot \exp[\frac{- B_0 \cdot \phi^{3/2} \cdot v(y)}{E}] 
\label{eq:4.1}
 \end{equation}
  
 In equation ~\ref{eq:4.1},  $A_0 = 1.54 \cdot 10^{-6} \frac{[A]}{[eV]}$  and $B_0 = -6.83  \frac{1}{[nm] \cdot \sqrt{[eV]}}$ are the universal Fowler constants derived from the triangular approximation model ~\citep{Forbes}; $j$ is the observed FN surface current density; $E$ is the enhanced field $E = \beta \cdot E_{wire}$, where $E_{wire}$ follows the cylindrical approximation model $E_{wire} = E_{ext} \cdot {\frac{p}{d}}$, where $\frac{p}{d}$ corresponds to the wire-pitch ($1$ mm) and wire diameter ($80 \mu$m) ratio ~\citep{princeton, edwards_eff}. In the equation for \textit{E}, $\beta$ is the field enhancement factor and represents the factor by which the applied field on the grids wires is enlarged on the localized protrusion tips.  Finally, $\phi$ corresponds to the total work potential of the emitter, and defined as $\phi = \phi_0 + V_0$ ~\citep{Bailey} with $\phi_0 = 4.71 $ eV equal to the work function of the Monel (a copper-nickel alloy) wires, and $V_0 = - 0.61$ eV equal to the work function of LXe. Note that in equation ~\ref{eq:4.1},  $v(y) \propto (1- y)\cdot(1 - y^2)^{1/2}$ and $t(y) \propto \frac{(1- y^2)^{1/2}}{(1 - y)^{1/2}} = \sqrt{( 1+ y)}$ are approximations of the elliptical integral Schottky- Nordheim functions, which are slowly varying functions of the enhanced field and work potential $\phi$, y = $(3.79 \cdot 10^{-5}) \cdot E^{1/2} \cdot \phi^{-1}$, and are close to a value of one ~\cite{Noer}. The FN current described by equation ~\ref{eq:4.1} is localized on the TPC grid protrusions, and the shape of these protrusions is described by the projection model with the emitter area $S$ ~\citep{Williams, smith, Cox}. 

According to the FN model, the electric field applied to the metallic gate grid surface causes the potential to acquire a finite width, resulting in the tunneling of conduction electrons near the Fermi level into the lower-potential region of the surrounding LXe \cite{Nordheim}. The field emission depends on multiple factors, such as the electric field strength, the thermal excitation of electrons above the Fermi level, and the energy distribution of the emitted electrons. Moreover, sharp points on the grid surface area, impurities embedded into the material and mechanical defects also enhance the electric fields, resulting in increased field emission \citep{grids, Bailey, Noer}. Field emission may play a role in decreasing single electron rate as the drift field grows (Fig.~\ref{fig:3}, Left). For a constant gate voltage and increasing drift field, as the number of field lines originating on the gate and ending on the anode should decrease, lowering the probability of gate emission electron collection, and thus reducing the overall number of single electrons extracted from the liquid. 

To determine the value of the FN emission current $j$, the enhancement factor $\beta$ must be calculated. The value of $\beta$ is found by fitting the equation ~\ref{eq:4.1} to the data and extracting $\beta$ from the slope, assuming that $E_{wire} = \frac{E}{\beta}$. We find $\beta = $ 1036 $\pm$ 52, and an area of the tip emitter of order $10^{-13}$ cm$^2$. This assumes a hemispherical projection height $ h \approx 1 \mu$m and $\beta = \frac{h}{d}$, where $d$ is the base diameter \citep{Noer, Brodie}. The enhancement factor $\beta$ grows as a function of the work function of the wires (Fig.~\ref{fig:4}, Right). The single electron rate is predicted from the Fowler-Nordheim expression ~\ref{eq:4.1}, and is of the same order of magnitude as the measured rate for varied extraction field (Fig.~\ref{fig:3}, Right). The error bars on the FN data points on Fig.~\ref{fig:4} (Left), are from the systematic errors resulting from the uncertainties of the extraction efficiency $\epsilon$ and the extraction field. We favor the interpretation of FN tunneling from metallic grids over electron tunneling from neutral impurities in the LXe for two reasons. First, the work function of neutral impurities is nearly twice as high as that of the metal surface ~\citep{Jingke, Keiko}; second, the electric field strength is typically much higher at the grid surface than in the bulk LXe.

There is evidence for simultaneous charge multiplication and photon production if the enhanced field is higher than 700 kV/cm \citep{grids, Bailey}; however, we did not observe any signs of breakdown currents. One possible explanation for that is the constrained combination of the enhancement factor $\beta$ and work function $\phi$, the value of $\phi$ is small, allowing a  lowered $\beta$. This tradeoff is quantified in Fig. ~\ref{fig:4} (Right), where a lower enhancement factor is compatible with a decreased work function of the wires, resulting in a increasing tunneling probability of the surface electrons with energies close to the Fermi level. The simple Jeffreys-Wentzel-Kramers-Brillouin (JWKB) approximation for the tunneling probability $D$ for an electron approaching a barrier of unreduced height $h_b$ of any well-determined barrier shape (where the transmission coefficient varies smoothly, decreases monotonically, as the energy at which an electron tunnels decreases) $D \approx \exp{(g_e \int M^{1/2} dz)} = \exp{(-\nu b h_b^{3/2}/ E)}$, where $\nu$ is a physical tunneling correction factor, $E$ is the constant barrier field, $g_e \approx 10.246$ $eV^{-1/2} nm^{-1}$ is the JWKB constant, $z$ is the distance between the emitter's surface, and $M$ is the motive energy function responsible for the shape of the barrier \citep{Forbes_2}. Since the FN current $j(E)$ in equation \ref{eq:4.1} is proportional to the tunneling probability $D$, comparing the exponential expressions, the work function $\phi$ is essentially the barrier height $h_b$. As the barrier height gets lower ($\phi$ is decreased), the area of the projection can be larger since the probability for electrons to tunnel through the barrier is increased, diminishing "the sharpness" of the projection tip. Since the area of the projection tip $A \approx 1/ \beta^2$, a wider tip corresponds to a lower enhancement factor value, which would be consistent with our observations Fig. ~\ref{fig:4} (Right) and also consistent with the fact that we do not observe breakdown electroluminescence. Alternatively, higher values of $\beta$ and $\phi$ may be accommodated by taking account of the stressed area, which is defined as the area on the cathode surface with electric field exceeding 90\% of the maximum electric field \citep{grids}. When the stressed area is minimized, breakdown currents are less likely to occur \citep{MicroBooNE,XeBrA}.

\subsection{Between the S1 and S2 pulses}
\begin{figure}[H]
% \begin{subfigure}{0.5\textwidth}
% \includegraphics[width=\linewidth]{pics/reg2_fg_new.pdf}
% \label{fig:5a}
% \caption{}
% \end{subfigure}
% \begin{subfigure}{0.5\textwidth}
% \includegraphics[width=\linewidth]{pics/reg2_fd_new.pdf}
% \label{fig:5b}
% \caption{}
% \end{subfigure}
\begin{subfigure}{0.5\textwidth}
\includegraphics[width=\linewidth]{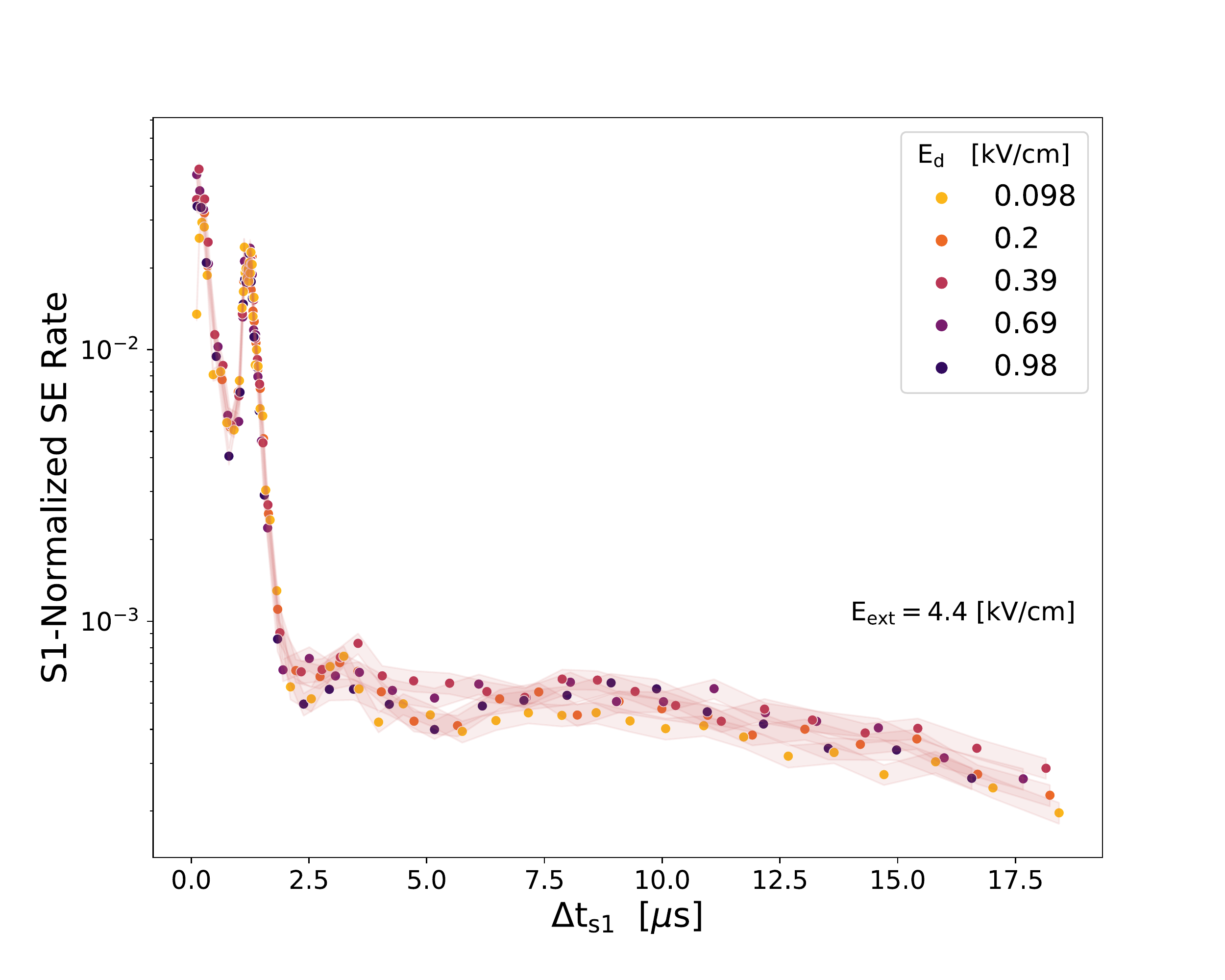}
\label{fig:5a}
\end{subfigure}
\begin{subfigure}{0.5\textwidth}
\includegraphics[width=\linewidth]{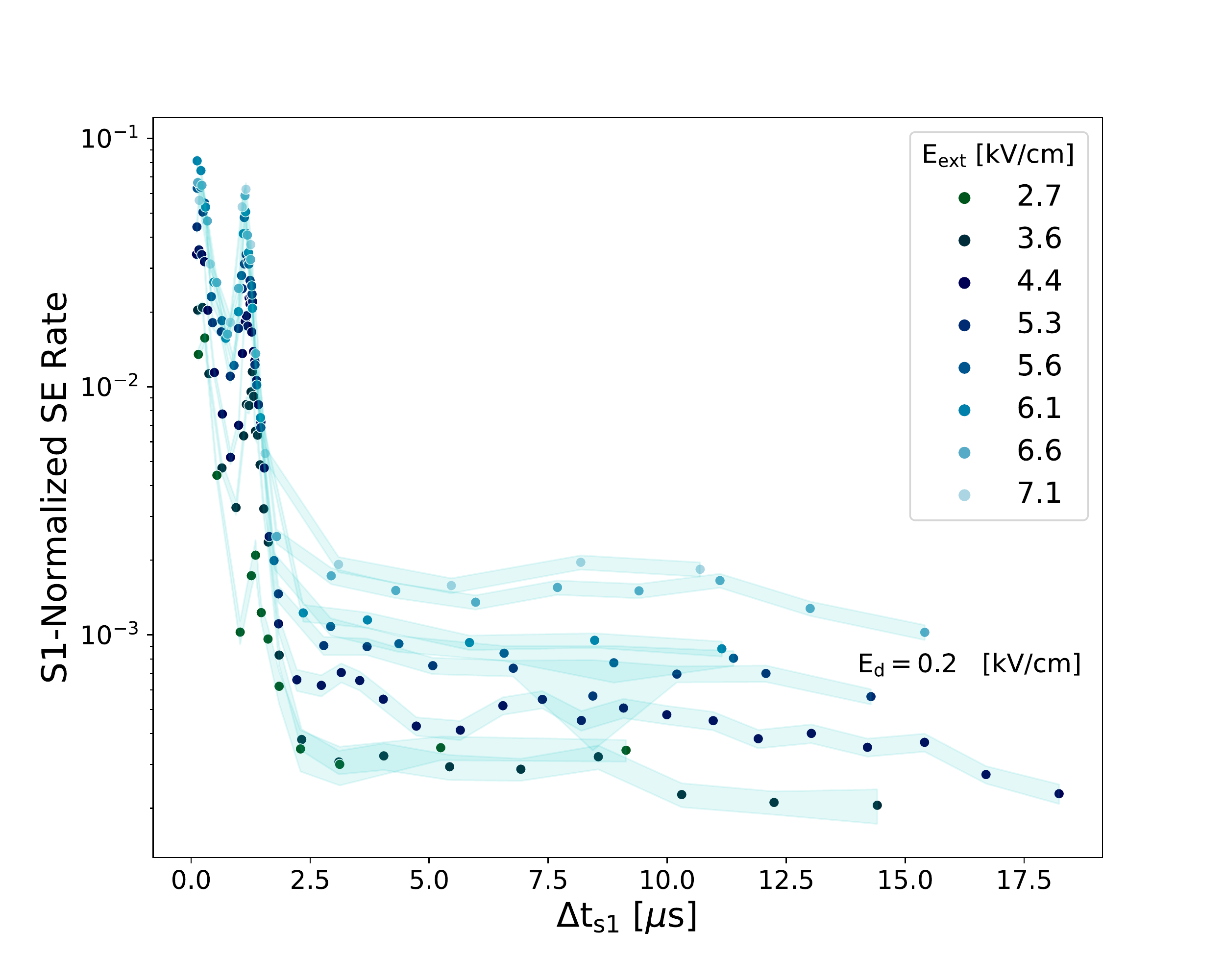}
\label{fig:5b}
\end{subfigure}

\caption{The single electron rate normalized by the S1 area for event drift times between 10 and 30 $\mu$s (a) as a function of the time elapsed relative to the S1 pulse $\Delta t_{s1}$ for a fixed extraction field at 4.4 kV/cm; (b) as a function of $\Delta t_{s1}$ for a fixed drift field of 0.2 kV/cm.}

\label{fig:5}

\end{figure}

In this section, we study the single electrons detected between the S1b and S2 pulses. The SE rate relative to the time of the S1 pulse is shown in Fig.~\ref{fig:5} for different drift and extraction fields. The SE rate is higher for larger extraction fields, increasing with detector's electron extraction efficiency (Fig. ~\ref{fig:5}, Right).  The spikes at 2~$\mu$s on Fig.~\ref{fig:5} are likely caused by S1 light causing the photoelectric effect on the gate grid ~\citep{pmt}. The SE rate is not observed to depend significantly on the event depth, determined from the drift time $\Delta t$.

\begin{figure}[H]
\includegraphics[width=0.48\linewidth]{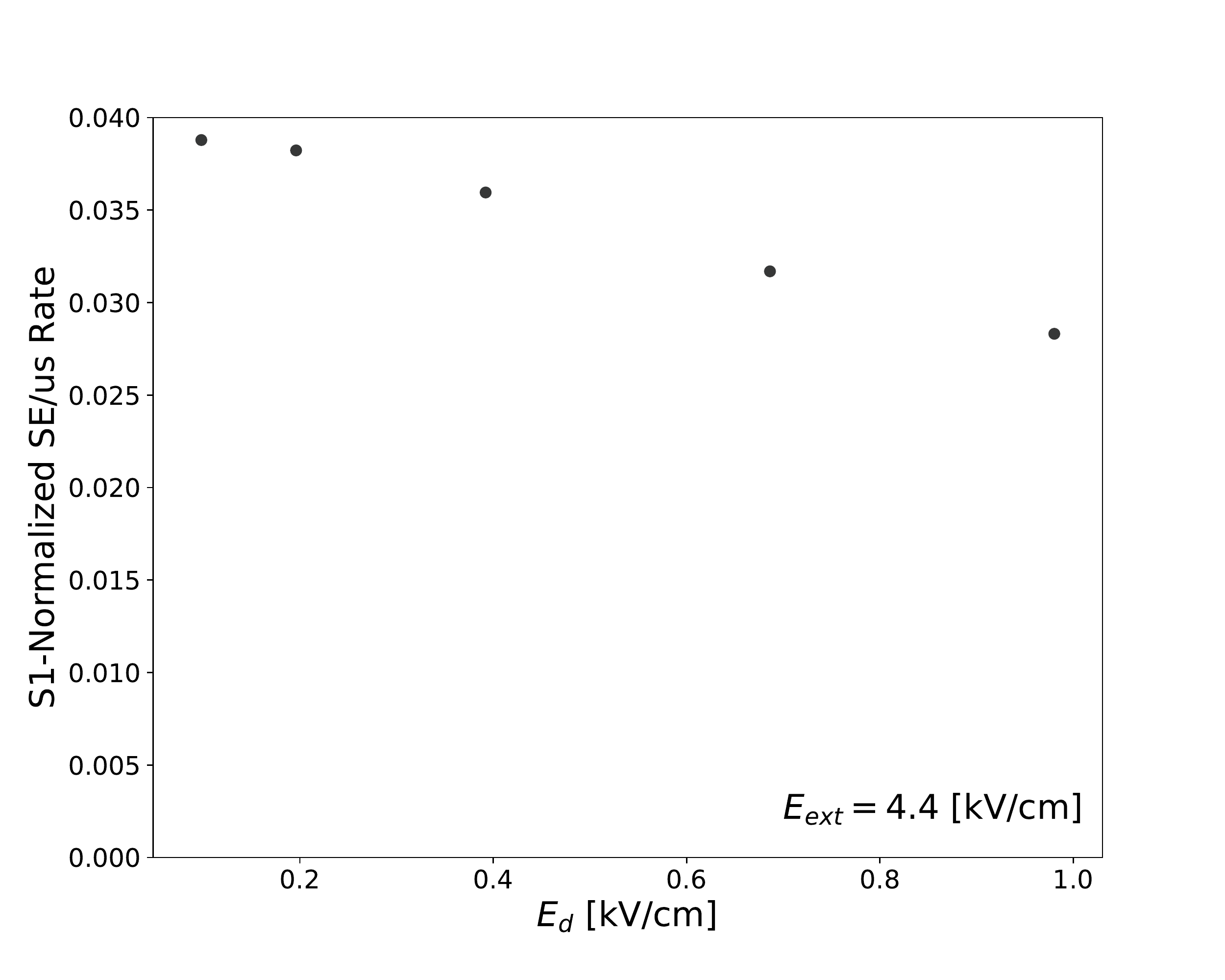} 
\includegraphics[width=0.48\linewidth]{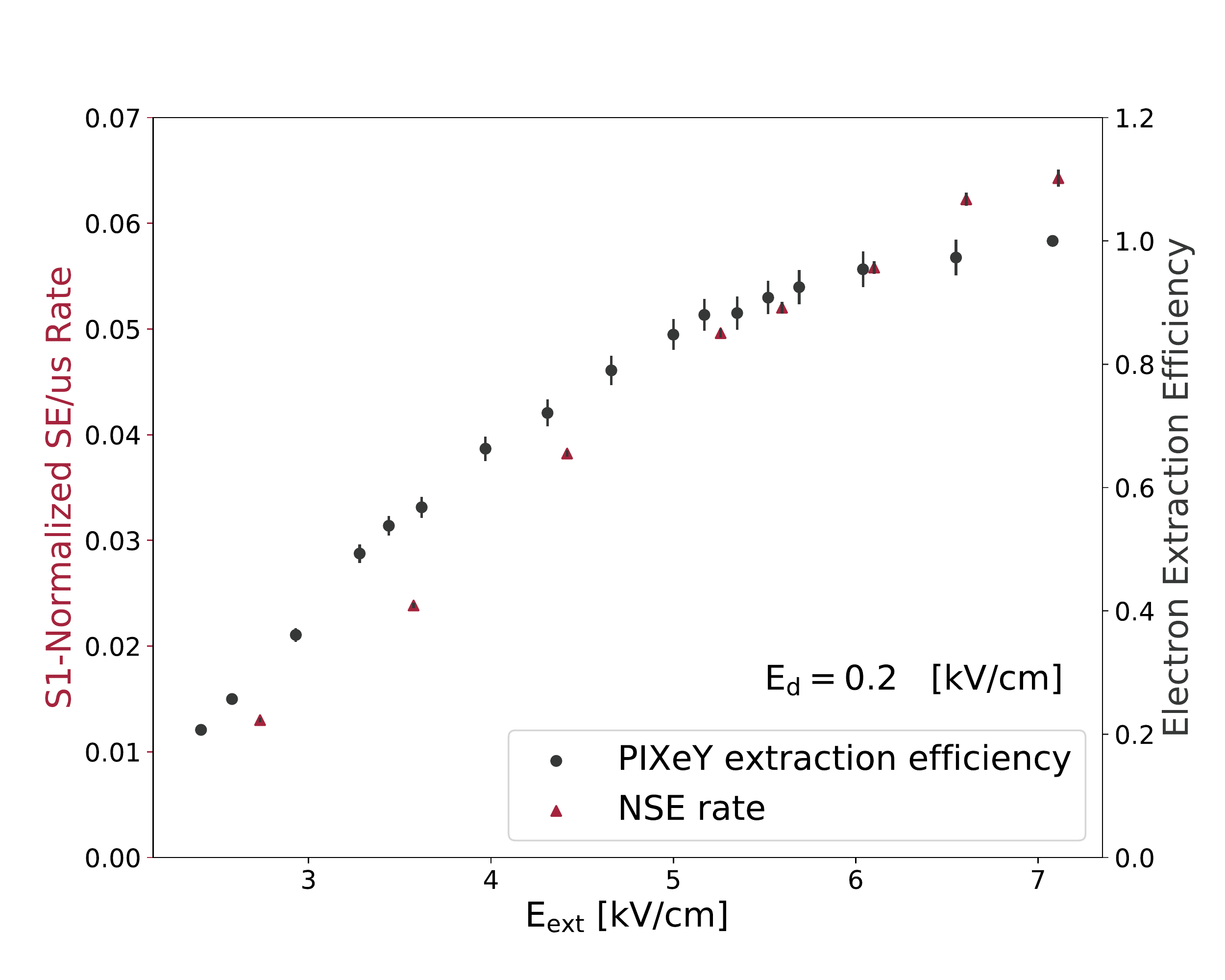}
\caption{The rate of single electrons is defined as an averaged sum of SE areas / S1 area per $\mu$s in between the S1 and S2 signals. \textit{Left:} for fixed extraction field $E_{ext} = 4.4$ kV/cm and varied drift field  $E_d$. One $\sigma$ error bars are shown. \textit{Right:} for fixed $E_d =  0.2$ kV/cm field and varied extraction field.}
\label{fig:6}
\end{figure}

The S1-normalized single electron background rate suppression for larger drift fields is not well understood (Fig.~\ref{fig:6}, Left). Under a drift field $E_d$, the electrons are not in thermal equilibrium with the liquid xenon, as they gain energy from the electric field. As these electrons are drifted to the extraction region, they meet impurities (of density $\rho$), and bind to them with energies of order 0.45 eV ~\cite{Aprile}. The formed impurities are later photo-ionized by the S1 and S2 light, resulting in single electron background after the S1 and S2 pulses, and causing the impurity density $\rho$ to be a function of the size of the signal size and rate in previous events. One possibility is simply that the higher electron drift velocity at high drift fields results in less interaction time between electrons and impurities, decreasing the chance that the electrons are captured, and thus decreasing the density of negatively-charged impurities that could later be photoionized.  Another hypothesis is that higher drift fields remove charged impurities more quickly, so their average density is lower for higher drift field.

Additionally, the SE background scales linearly with the extraction field (Fig.~\ref{fig:6}, Right) showing that the single electrons are either extracted on their first attempt, or not at all on this time scale. Indeed, once $E_d$ is constant, the photoionization cross-section is fixed for that particular drift field value. As a result, the single electron rate from photoionization remains constant, so these electrons will keep drifting under constant drift field to the liquid surface and be extracted with an efficiency determined by the extraction field $E_{ext}$. Photoionization of the impurities by S1 light dominates over FN emission in this region.

There is a possibility that the distribution of negatively charged impurities is not constant. Under the assumption of good purity, almost every free electron produced in the liquid bulk reaches the liquid surface. However the free electrons are born at different depths. If assuming that liquid motion keeps the number of neutral impurities uniform and if that negative ion formation rate is proportional to the neutral impurity density and the flux of free electrons, then there will be a gradient in negative ion density, with the negative ions more highly concentrated toward the top of the detector. This hypothesis could be partially responsible for the downward slopes seen in figures (Fig.~\ref{fig:5}) and (Fig.~\ref{fig:7}). In this case, the SE rate would be an indirect consequence of the total light exposure over a much longer timescale.

\subsection{After the S2 Signal}
\subsubsection{Between S2 and S2 + Maximum Drift Time}
\begin{figure}[H]

\includegraphics[width=0.5\linewidth,  ]{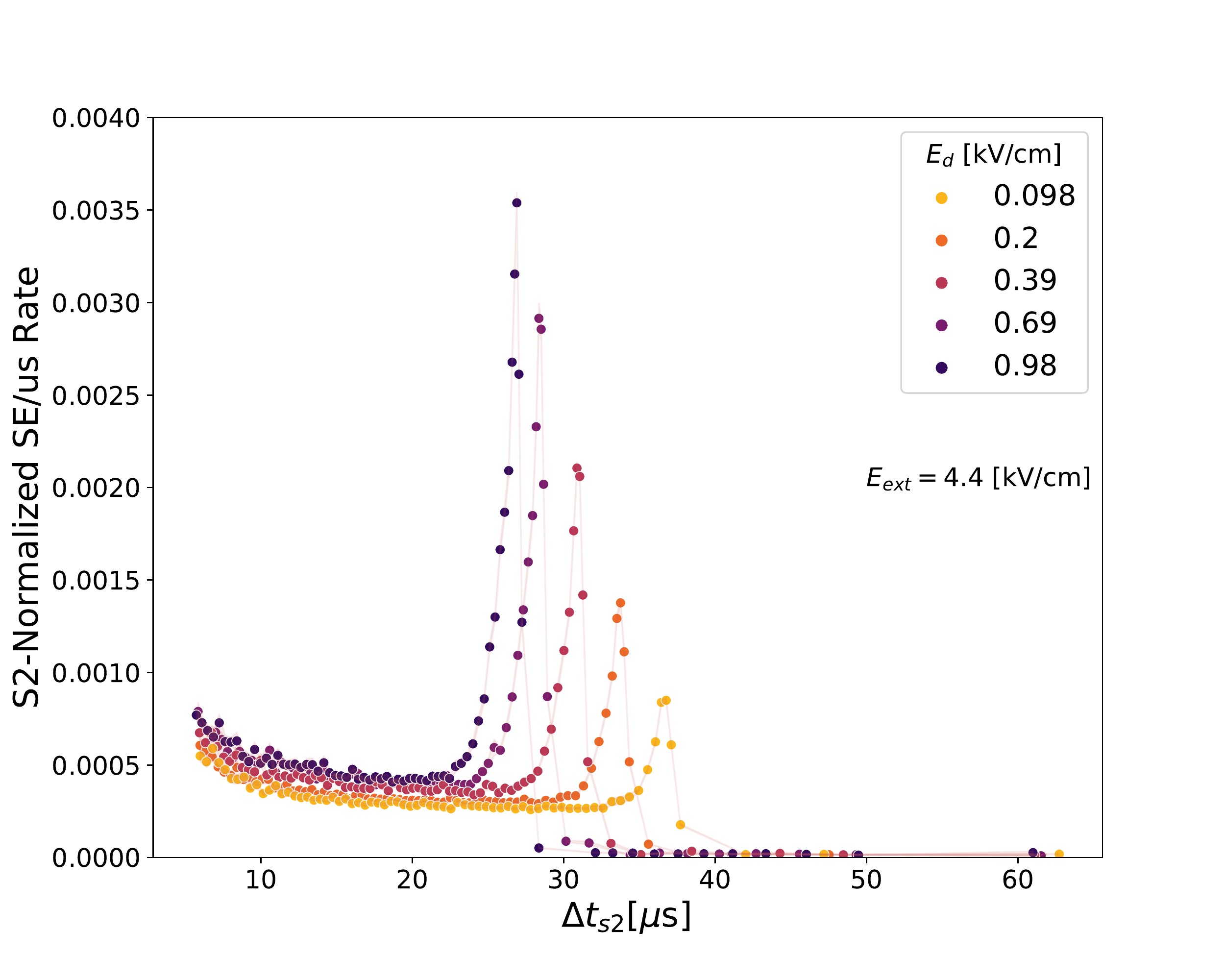} 
\includegraphics[width=0.5\linewidth,  ]{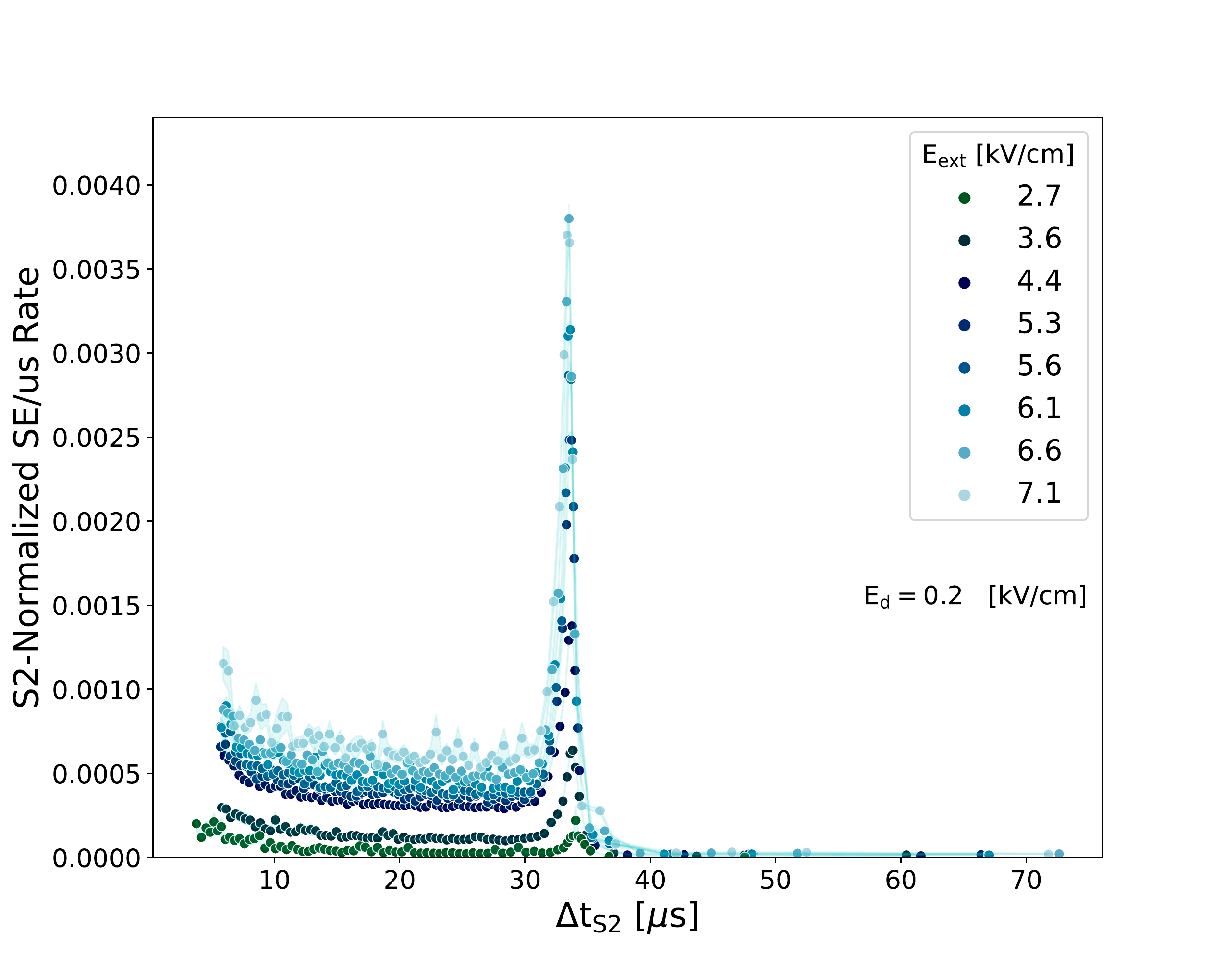}

\caption{The normalized rate of single electrons is defined as the sum of SE areas / S2 area, per $\mu s$ after the S2 signal. The cathode spike is peaked around the maximum drift time. \textit{ Left:}  The extraction field is fixed at 4.4 kV/cm, and the drift field varied. \textit{ Right:} The drift field is constant at 0.2 kV/cm, while the extraction field is varied.}
\label{fig:7}
\end{figure}

The single electron background is calculated between the end of the S2 and S2 + $t_{max}$, where $t_{max}$ corresponds to the maximum drift time of a single electron, and is approximately 35 $\mu s$ relative to the S2 (Fig. ~\ref{fig:7}).

\begin{figure}[H]
\includegraphics[width=0.48\linewidth  ]{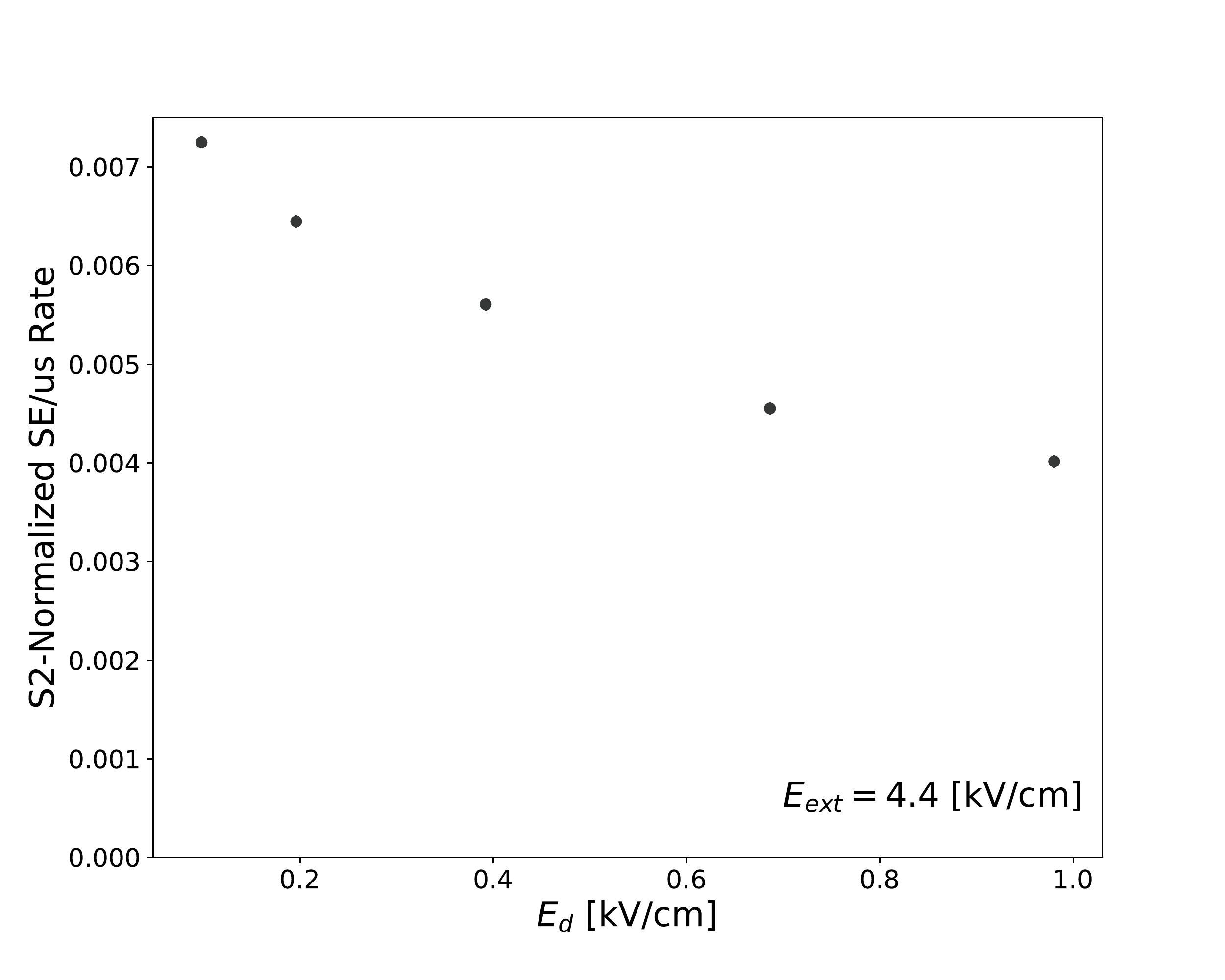} 
\includegraphics[width=0.48\linewidth  ]{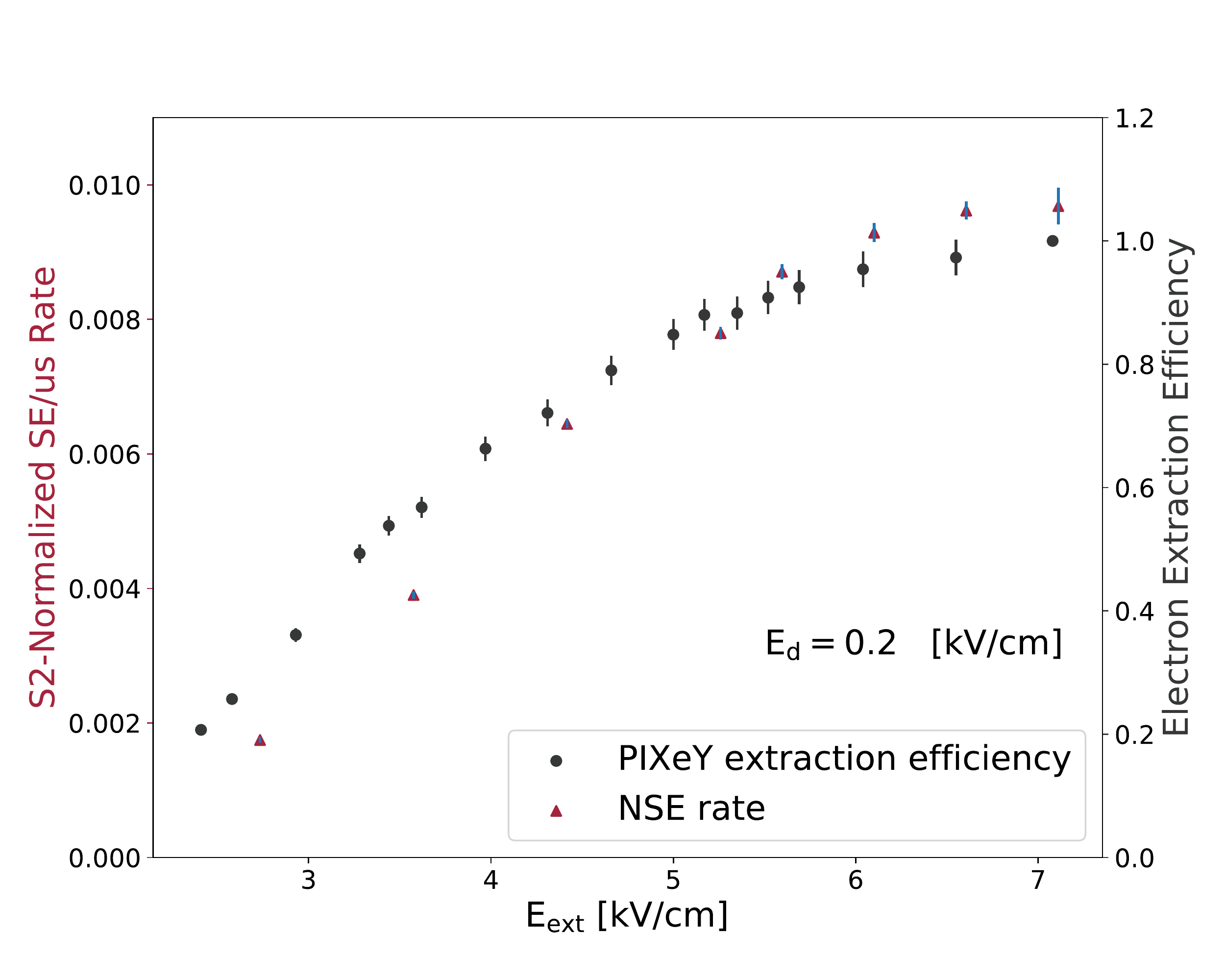}
\caption{Number of single electrons between S2 and S2 + $t_{max}$ normalized with S2 area, per $\mu s$.  \textit{Left:} SE decreases with drift field for fixed 4.4 kV/cm extraction field. \textit{Right:} Normalized SE rate for fixed drift field 0.2 kV/cm and varied extraction field. The shape of the SE rate is compatible with the PIXeY extraction efficiency curve.}
\label{fig:8}
\end{figure}

The S2-normalized SE rate also declines as a function of the drift field (Fig. ~\ref{fig:8}, Left), in a similar way that it does in between the S1 and S2 pulses. There is an additional photoelectric effect on the TPC grids and the impurities $\rho$ by the S2 light, since this emission rate is mainly dependent on the size of the total light signal. 
Again, the SE rate trend (Fig. ~\ref{fig:8}, Right) closely follows the PIXeY extraction efficiency curve indicating that the photoionization effect dominates over FN emission, similar to what has been observed in the region between S1 and S2 pulses.

\subsubsection{The Cathode Spike and Quantum Efficiency}

\begin{figure}[H]
\begin{subfigure}{0.5\textwidth}
\includegraphics[width=1.1\linewidth]{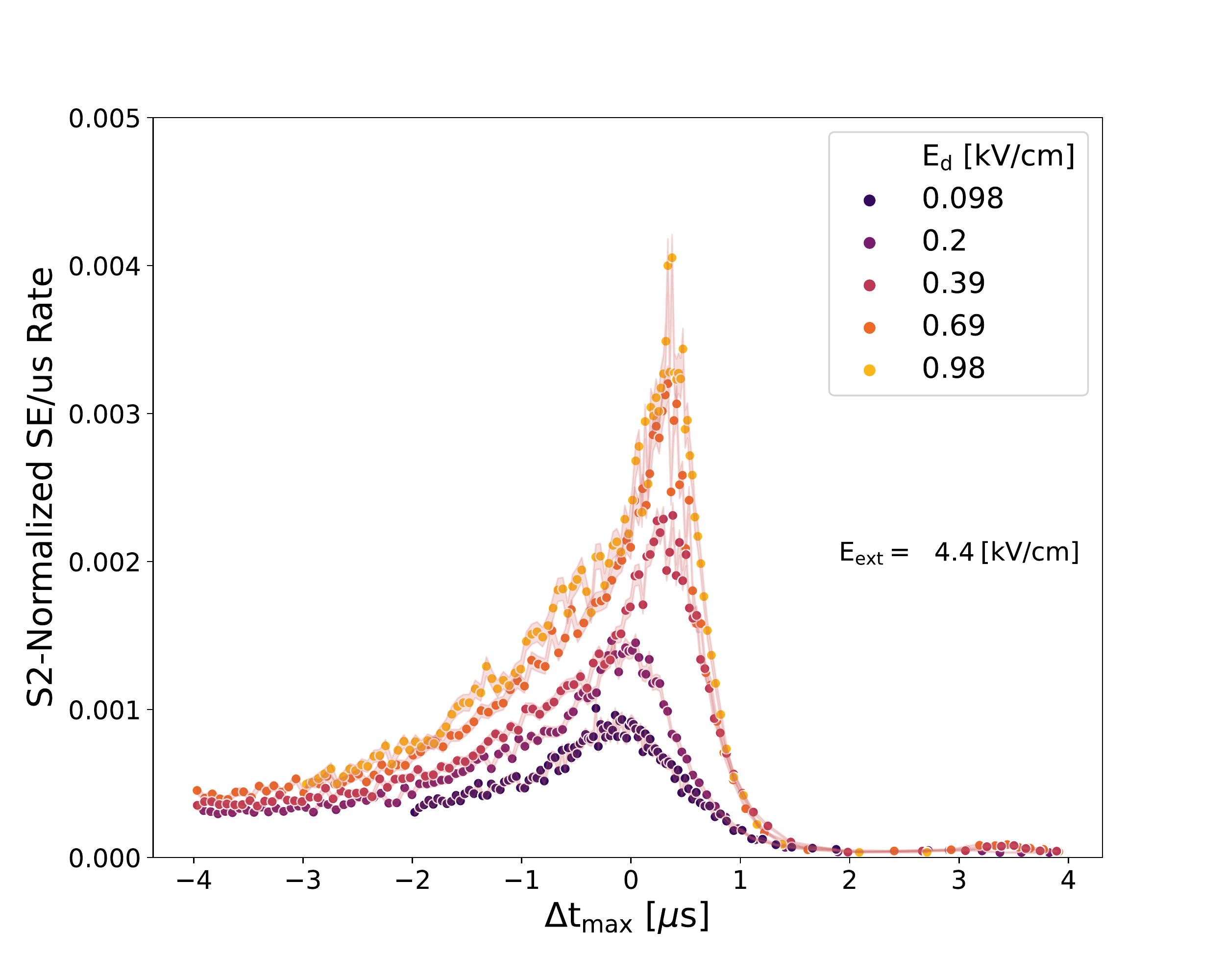}
\end{subfigure}
\begin{subfigure}{0.5\textwidth}
\includegraphics[width=1.1\linewidth]{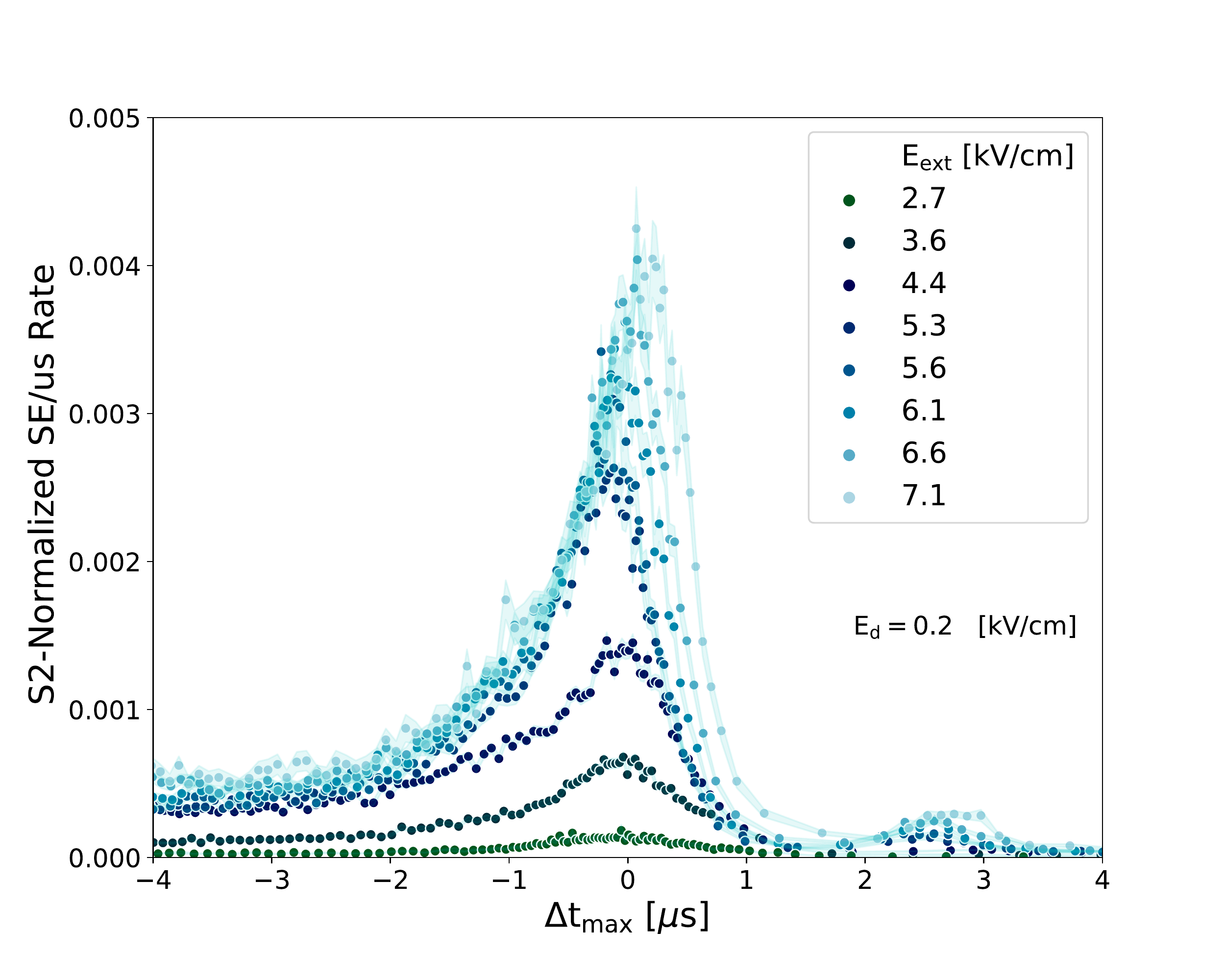}
\end{subfigure}
\caption{The cathode spike single electrons rate normalized with the S2 signal area. $\Delta t_{max}$ is a single electron time drift with respect to S2 time + max available drift time. To center all the peaks around zero, the maximum available drift time of SE was subtracted from time between S1 and S2 pulses. \textit{ Left}: The SE rate in the cathode spike region for fixed extraction field at 5.89 kV/cm and varied drift field.\textit{ Right:} The SE rate for fixed drift field at 0.2 kV/cm and varied extraction field. }
\label{fig:9}
\end{figure}

The cathode spike is the timescale region near the maximum drift time of the single electrons produced by the S2 pulse  (Fig. ~\ref{fig:7} and Fig. ~\ref{fig:9}).

Although the drift fields are not strong enough to produce significant Fowler-Nordheim emission from the cathode grids, they can still lower the Schottky potential to enhance the photoelectric effect. As a result, we observe drift field-dependent behaviour of the single electron rate (Fig.~\ref{fig:9}, Left). Additionally we observe two spikes growing as a function of the extraction field on (Fig.~\ref{fig:9}, Right), where the tallest spike corresponds to the cathode grids photoelectric effect from the S2 VUV light, and the short peak after $2 \mu$s for the field values above $E_{ext} = 5.3$ kV/cm is an echo due to secondary electron emission from the gate grid. 

The large spike from the photoelectric effect on (Fig.~\ref{fig:9}, Right) grows for bigger extraction field because of the increased SE/S2 size, and increased electron extraction efficiency. The origin of the gate echo single electrons is similar; however the Schottky suppression is higher comparing to the cathodic grids, because of the larger extraction fields. So, the gate photocurrent is mainly produced by both FN emission and photoelectric effect caused by single electron electroluminescence (Fowler hypothesis). A fraction of electrons striking the surface enhance their normal energy by $\hbar \nu$ on account of the absorption of a photon. Once they energized by light, the probability for the grid electrons of tunneling through the potential energy barrier adjacent to the negatively charged surface is higher. Hence it is possible to generate the field emission  by the incidence of light enhancement, even when the frequency is below the photoelectric threshold frequency \citep{Sodha, F-hypothesis}.

% The SE rate due to the FN effect, and the cathode spike from the photoelectric effect are theoretically related, both growing at higher electric fields. Inspired by studies on quantum model for photoemission ~\cite{Zhou}, the time-dependent surface potential barrier $\Phi(x, t)$ can be written as a function of the Fermi level of the metal $E_f$, the effective work function $W$ which includes the potential barrier lowering by the electric field $E$, effective metal work function \textit{$W_0$}, the electron charge $e$, and the permittivity of the xenon $\epsilon_{Xe}$; $F_1$ is the magnitude of the light field (in analogy with laser fields from ~\cite{Zhou}) with frequency $\omega$: $\Phi(x, t) =  E_f + W(E) - e\cdot F_1 \cdot x \cos(\omega t), \text{for} x\geq 0\\$ with x representing the position of metal-xenon interface.

Following the argument in ~\cite{Zhou} with the  potential given for the case of non-zero electric field, the resulting equation for quantum efficiency (QE), the number of electrons produced from S2 light photoionizing the cathode wire grids, is predicted to be a function of applied electric field, temperature, and incident light frequency. It is predicted that QE grows for larger electric fields, which explains why the cathode spike is taller for larger drift fields on Fig.~\ref{fig:9}. These observations motivate attempts to unify photoemission, possibly with Fowler-Nordheim theory as a function of Schottky potential, electric field, temperature, and light frequency, though this is beyond the scope of this paper.

The cathode spikes (Fig. ~\ref{fig:9}) can be used to study the cathode quantum efficiency, which is defined as the number of electrons ($N_e$) produced as a result of light hitting the wire grids on the cathode ($N_{ph}$). 

\begin{equation*}
    QE = \frac{N_e}{N_{ph}},
\end{equation*}

The number of photons $N_{ph}$ is proportional to photon pulse size and also a function of geometrical factors, such as the coefficient of visibility (how many of the wires are "visible" by the light), the $z$ dependent solid angle, and probability that the light will hit the grid wires. Consider photons ${N_{ph}}$ traveling along the z axis of the fiducial volume and radius r = 6.5 cm, which is the maximum radius available for the electrons distributed in x-y plane, 2.55 cm away from the cathode wires with 80 $\mu$m diameter and average length of 0.138 m. These photons will hit the grids with a certain probability and knock off the electrons from the cathode grids $N_e$. Such a scenario causes only 4$\%$  of the grid wires given our detector geometry, to be visible by the S2 light. Note that the QE includes the effect of the fraction of field lines on the cathodic wires going up. In fact, some electrons emitted straight up will be carried upwards by the field lines. However, electrons emitted from the side of the wire might follow field lines that end up going downward. The relative area of the wire that has field lines going up will change with the drift field.

\begin{figure}[H]
\begin{center}
\includegraphics[width=0.60\linewidth]{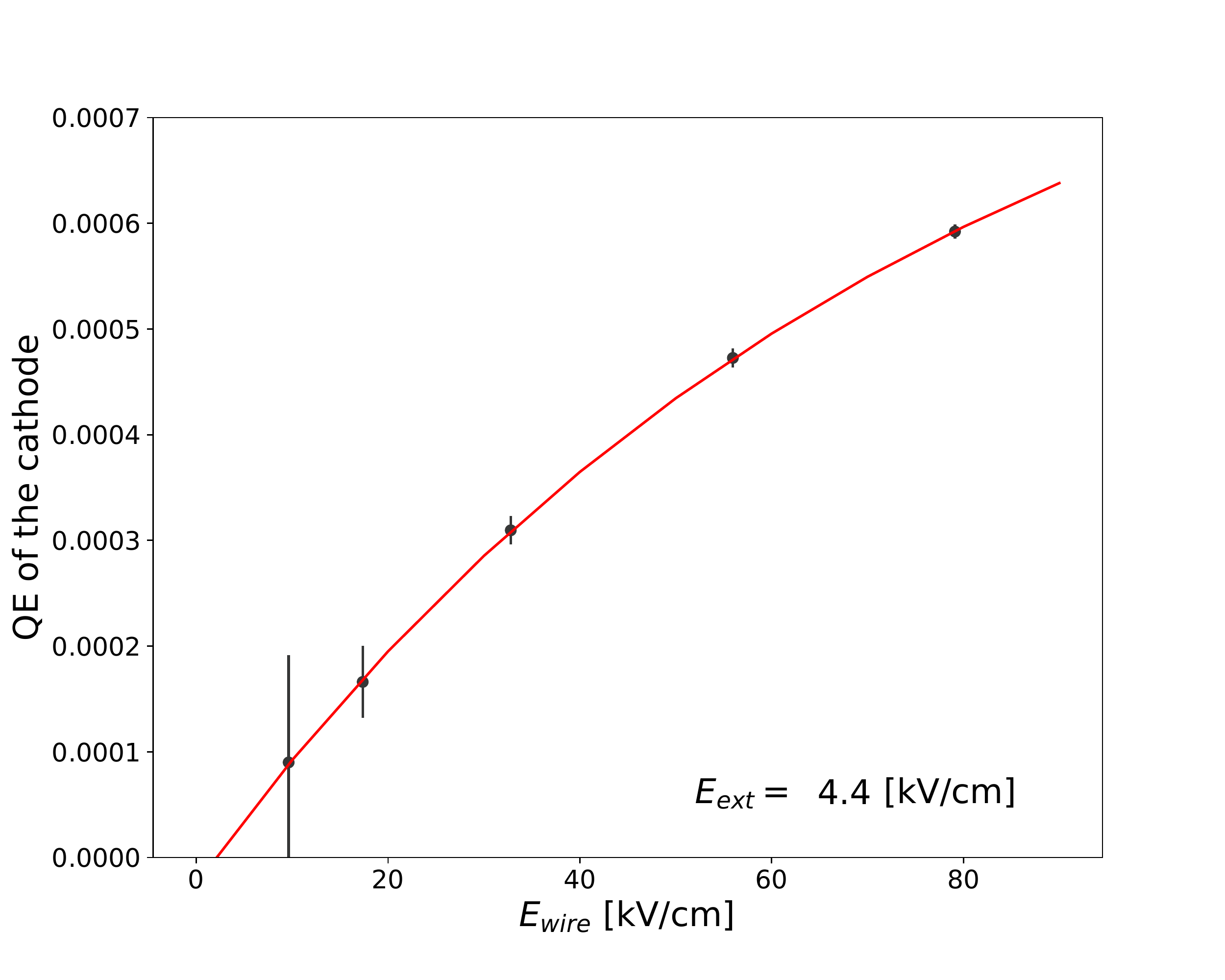} 
\end{center}

\caption{Quantum efficiency of the cathode wires as a function of drift electric fields on the TPC wires, assuming cylindrical approximation, with the exponential function fit. Quantum efficiency is defined as the amount of electrons produced as a result of S2 light photoionizing the cathode wire grids. The error bars are dominated by the systematic uncertainty of the electron extraction efficiency, for fixed extraction field at 4.4 kV/cm. } 
\label{fig:10}
\end{figure}

The number of produced electrons $N_e$ for $\Delta t_{max}$ between $-1 \mu s$ and $1 \mu s$ is proportional to S2 light (Fig. ~\ref{fig:9}), the QE is growing as a function in the case of both drift and QE is flat near $2 \cdot 10^{-4}$ for fixed drift field at 0.2 kV/cm. The quantum efficiency is fit to an exponential function $c - A \cdot \exp{\frac{-E_{wire}}{E_c}}$, where $c = 9 \cdot 10^{-4}$, $A = 9.6 10^{-4}$ is the maximum quantum efficiency, and some characteristic field $E_c = 80$ kV/cm
(Fig. ~\ref{fig:10}). The physical meaning of the characteristic field and $c$ constant is unclear; more studies should be done, employing the extended Fowler-Nordheim theory with the photoelectric effect. Quantum efficiency is also seen to grow with field in studies of CsI immersed in LXe \citep{QE}.

\subsubsection{After The Cathode Spike}

\begin{figure}[H]
\includegraphics[width=0.48\linewidth]{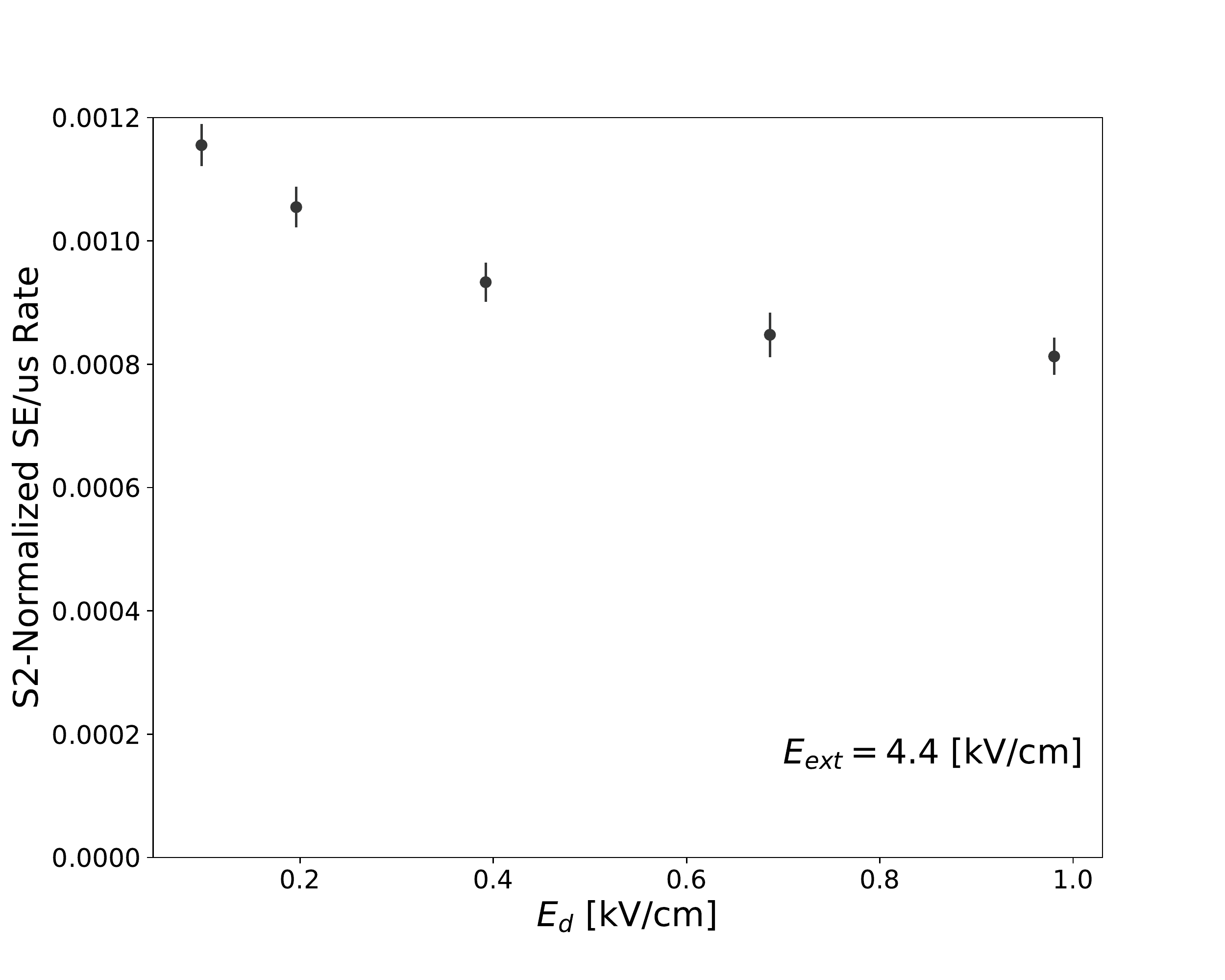} 
\includegraphics[width=0.48\linewidth]{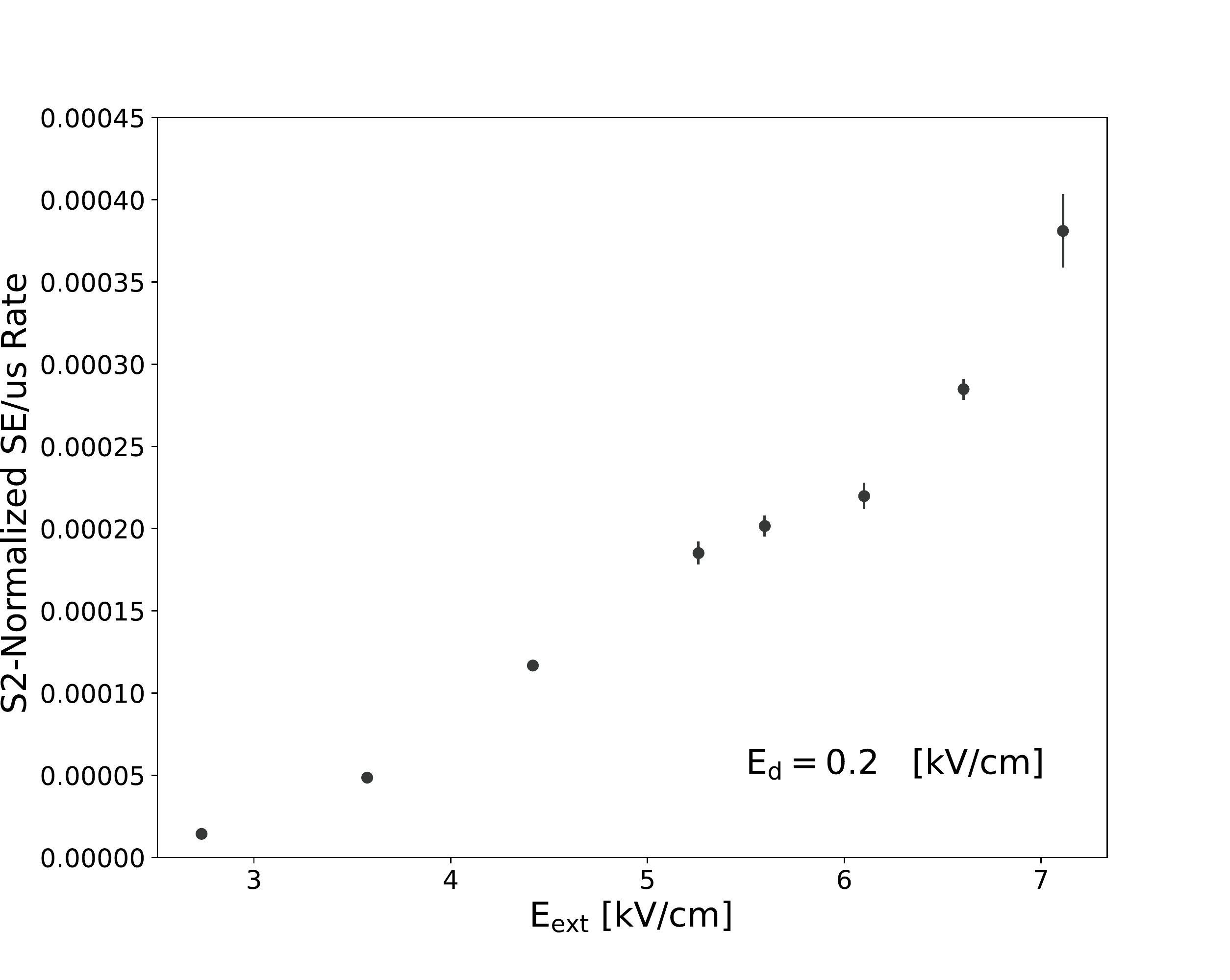}

\caption{The rate of single electrons normalized with S2 area after the cathode spike. \textit{Left:} for fixed extraction field 4.4 kV/cm and varied drift field. \textit{Right:} for fixed drift field 0.2 kV/cm and varied extraction field.}
\label{fig:11}
\end{figure}

The single electrons after the cathode spike behave similarly as in previous sections: the rate increases with growing extraction field, and diminishes with higher drift field (Fig. ~\ref{fig:11}). However, there is a major distinction in the origin of these single electrons. Considering the SE rate as a function of drift field in LXe (Fig. ~\ref{fig:11} and Fig. ~\ref{fig:3} left panels), general background single electrons are pure artifact of the Fowler-Nordheim emission, while the single electrons after the cathode spike (Fig. ~\ref{fig:11}, left) originate from both field emission and spontaneous release of the individual single electrons by LXe impurities, which were formed in the previous event.

\section{Discussion}

It has been observed that there are two main origins of spurious single electrons: the TPC grids (cathode, gate grids) and impurities in LXe. Moreover, there are independent mechanisms which cause these sources to produce single electrons: photoionization by the S1 and S2 light as well as field emission. The TPC grids are a source of photo-current from the S1 and S2 light; and grid protrusions enhance the Fowler-Nordheim emission current. 

On all observed time scales, the single electron rate increases with higher extraction field, because of both increased FN current and higher electron extraction efficiency.  We find a few possible explanations for the declining single electron rate with higher drift fields, which are all summarized in equation (5.1). In this equation, $N$ is the total rate of single electrons observed and is linear with $S$, the signal size, to leading order, $G$ is the characteristic length scale, representing the average distance traversed by a photon in the bulk, $\rho_n$ is the density of neutral impurities with photoionization cross-section $\sigma_n (E)$, $\rho_-$ is the density of negative impurities with photoionization cross-section $\sigma_- (E)$, $\epsilon_{ext}$ is the electron extraction efficiency, and $N_{FN}$ is the rate of single electrons produced from Fowler-Nordheim emission.
\begin{equation}
 N \propto N_{FN}\epsilon_{ext} + \epsilon_{ext} \cdot S \cdot  G \cdot (\rho_n \cdot \sigma_n (E) +  \rho_- \cdot \sigma_- (E) ) 
 \label{5.1}
\end{equation}

The first hypothesis is that the LXe impurities have electron capture cross-sections $\sigma_c (E)$ that decrease with electric field, thus decreasing the density of the impurities. The second hypothesis is that the higher drift field causes the charged impurities to be drifted out of the detector more quickly, thus diminishing this source of single electron background. However, note that the faster-drifting electrons have less time in contact with impurities and this would also decrease the single electron rate for higher drift fields. It is also possible that the photoionization cross-section $\sigma (E)$ of neutral or negatively charged impurities can be field-dependent, possibly decreasing with growing drift field.

Note that photoionization of $\rho_n$ and $\rho_-$ are two different physical mechanisms. In a case of the negatively charged impurities, the photoionization cross-section is likely higher, since the ionization work function of these impurities is below 1 eV. For the higher-density neutral impurities, the ionization energy is higher than the S1 or S2 light can provide, but photoionization is still possible because of atomic thermal energy gain in the electric field \citep{Jingke, Keiko}.
 
Exploring different limits of the equation ~\eqref{5.1}, we can explain the behavior of single electrons on all timescale regions relative to the S1 and S2 pulses. Considering the region before the S1a pulse, the observed rate $N$ is dominated by the product of the FN emission and the extraction efficiency $N \propto N_{FN}\epsilon_{ext}$. For times after the S1 pulse the observed rate is dominated by photoionization-induced backgrounds, which decrease with increasing $E_d$ but increase with increasing $E_{ext}$. In this regime, photoionization dominates over field emission. 

The origin of single electrons still warrants further research. As experiments like LZ, XENONnT, and PANDAX-4T come online, they will provide valuable insight on this crucial background. Our results are consistent with previous analyses (i.e., those by LUX, XENON, ZEPLIN, and Sorensen and Kamdin), but this analysis is the first to approach the question from a perspective of Fowler-Nordheim (FN) theories. Future discussions on the single electron background should also consider this perspective, in particular regarding single electron emission from the cathode and gate grids.

\section{Conclusion}

With the PIXeY TPC, we studied the behavior of the single electrons for a variety of drift and extraction fields, and for several timescale regions relative to the S1 and S2 pulses from \ce{^{83m}_{}Kr} calibration events. The TPC grids and LXe impurities are the main sources of single electrons, and the electron emission is stimulated by both the Fowler-Nordheim effect and the photoelectric effect from S1 and S2 VUV photons. The field emission from the TPC grids is accurately described by the Fowler-Nordheim theory. The total electron emission rate is a function of multiple variables: the detector geometry, extraction efficiency, the density of impurities, and the light signal size.

\section{Acknowledgments}
We thank Jingke Xu and Henrique Araujo for helpful discussions. 
We are grateful for support from DHS grant 2011-DN-007-ARI056-02, NSF grant PHY-1312561, and DOE grant DE-FG02-94ER40870.  The $^{83}$Rb used in this research to produce $\ce{^{83m}_{}Kr}$ was supplied by the United States Department of Energy Office of Science by the Isotope Program in the Office of Nuclear Physics.

\bibliographystyle{plain}

\input{bib.tex}

\end{document}

%% file: bib.tex
%%%%% CLEAR DOUBLE PAGE!
\newpage{\pagestyle{empty}\cleardoublepage}